\documentclass[11pt,a4paper]{article}
\pdfoutput=1
\usepackage{jcappub}

\usepackage{amsmath,amssymb,amsfonts}
\usepackage{graphics,graphicx}
\usepackage{color}

\graphicspath{{figs/},{figs/visit/}}

\newcommand{\figref}[1]{figure~\ref{#1}}
\newcommand{\Figref}[1]{Figure~\ref{#1}}

\def\ie{i.e.}

\def\laplacian{Laplacian}

\newcommand{\twofieldinflaton}{\url{www.cita.utoronto.ca/~jbraden/Movies/twofield_inflaton.avi}}
\newcommand{\twofieldtunnel}{\url{www.cita.utoronto.ca/~jbraden/Movies/twofield_tunnel.avi}}
\newcommand{\plateaufield}{\url{www.cita.utoronto.ca/~jbraden/Movies/collision_plateau_del0.1_field.avi}}
\newcommand{\thinwallexactfield}{\url{www.cita.utoronto.ca/~jbraden/Movies/linear_del0.1_exactprofile_field.avi}}  
\newcommand{\thinwallrho}{\url{www.cita.utoronto.ca/~jbraden/Movies/linear_del0.1_rhocontours.mp4}}  
\newcommand{\thinwallapproxfield}{\url{www.cita.utoronto.ca/~jbraden/Movies/linear_del0.1_thinwall_field.avi}}
\newcommand{\thinwallflucfield}{\url{www.cita.utoronto.ca/~jbraden/Movies/linear_del0.1_wfluc_field.avi}}  
\newcommand{\thickwallfield}{\url{www.cita.utoronto.ca/~jbraden/Movies/thick_wall_collision.avi}}

\begin{document}

\title{Cosmic bubble and domain wall instabilities III: the role of oscillons in three-dimensional bubble collisions}
\author[a]{J. Richard Bond,}
\author[a,b,c]{Jonathan Braden}
\author[d]{and Laura Mersini-Houghton}

\affiliation[a]{CITA, University of Toronto, 60 St. George Street, Toronto, ON, M5S 3H8, Canada}
\affiliation[b]{Department of Physics, University of Toronto, 60 St. George Street, Toronto, ON, M5S 3H8, Canada}
\affiliation[c]{Department of Physics and Astronomy, University College London, London, WC1E 6BT, U.K.}
\affiliation[d]{Department of Physics and Astronomy, University of North Carolina-Chapel Hill, NC 27599-3255, USA}

\emailAdd{bond@cita.utoronto.ca}
\emailAdd{j.braden@ucl.ac.uk}
\emailAdd{mersini@physics.unc.edu}

\abstract{We study collisions between pairs of bubbles nucleated in an ambient false vacuum.  For the first time, we include the effects of small initial (quantum) fluctuations around the instanton profiles describing the most likely initial bubble profile.
Past studies of this problem neglect these fluctuations and work under the assumption that the collisions posess an exact SO(2,1) symmetry.
We use three-dimensional lattice simulations to demonstrate that for double-well potentials, small initial perturbations to this symmetry can be amplified as the system evolves.
Initially the amplification is well-described by linear perturbation theory around the SO(2,1) background, 
but the onset of strong nonlinearities amongst the fluctuations quickly leads to a drastic breaking of the original SO(2,1) symmetry and the production of oscillons in the collision region.
We explore several single-field models, and we find it is hard to both realize inflation inside of a bubble and produce oscillons in a collision.
Finally, we extend our results to a simple two-field model.
The additional freedom allowed by the second field allows us to construct viable inflationary models that allow oscillon production in collisions.
The breaking of the SO(2,1) symmetry allows for a new class of observational signatures from bubble collisions that do not posess azimuthal symmetry,
including the production of gravitational waves which cannot be supported by an SO(2,1) spacetime.
}

\date{\today}
\maketitle

\section{Introduction}
This is the third installment in our investigation of fluctuations around highly symmetric domain wall collisions.
We began with a linearized analysis in~\cite{ref:bbm1} and continued with numerical lattice simulations of colliding nearly planar walls in~\cite{ref:bbm2}.
In this paper our focus is on the full nonlinear three-dimensional dynamics of collisions between pairs of bubbles nucleating in an ambient false vacuum.
Such collisions are expected to occur in inflationary models where our observable universe is contained within one of these nucleated bubbles.
Among current cosmological paradigms, this class of models are often discussed in the context of the multiverse, and more specifically the landscape~\cite{Bousso:2000xa,Susskind:2003kw}.
We explain more precisely our notion of the landscape and the embedding of inflationary bubble models below.
Bubble collisions are also relevant to the study of first-order phase transitions occurring in both particle and condensed matter physics.
In the cosmological context, the universe may have undergone a number of these transitions in its infancy. 
Of course, collisions between more than two bubbles are required for the phase transition to complete, although individual collisions between pairs of bubbles also contribute.

Collisions between \emph{pairs of bubbles} are most frequently considered in scenarios of open inflation following false vacuum decay.
In this class of models, our local cosmological history proceeds as follows.
A bubble nucleates within a surrounding false vacuum corresponding to the Big Bang ``singularity'' of our observable universe.
The interior of the bubble undergoes $\Delta\ln a \gtrsim 50$ efolds of inflationary expansion to dilute the effects of spatial curvature and seed the density perturbations that will eventually collapse to form the structure we see in the universe today.
Inflation inside the bubble ends and the universe (p)reheats, thus starting the standard thermal history of the universe.
Meanwhile the scalar fields relax to a new deSitter false vacuum with $\rho_\Lambda \sim 10^{-120}M_P^4$ in order to explain the present accelerated expansion of the universe.
Of course, since our universe arises from the nucleation of a bubble in this scenario, additional bubbles can nucleate in the ultra-large scale ambient false vacuum as well.
This presents the interesting possibility of a collision between the bubble containing our observable universe and another bubble.

The scenario described above is often invoked in the context of the landscape, and we now outline the prevailing picture for this embedding.
Motivated by current ideas in high-energy theory, it is postulated that at the energy scales relevant for inflation there are many effective degrees of freedom beyond those of the Standard Model.
For simplicity, consider the case where these degrees of freedom can be modelled as a large number of scalar quantum fields interacting through a potential and semiclassical general relativity.\footnote{This neglects various gauge and fermionic fields, as well as discrete degrees of freedom such as units of flux.  However, it is sufficient to illustrate the setup.}
We also assume some mechanism exists to seed random initial conditions on the potential surface at different points in space.\footnote{As an example of such a seeding mechanism, some portion of the wavefunction for the (unknown) fundamental degrees of freedom may describe QFT coupled to semiclassical general relativity.  If this portion of the wavefunction decohered, we would naturally expect it to seed stochastic initial conditions for the emergent scalar fields.  Dynamics on the potential surface (such as eternal inflation) may also serve to populate the landscape.}
The combination of a large number of dynamical scalar fields interacting through a potential with a mechanism to seed random initial conditions is what we will call the \emph{landscape paradigm}.  
The fields then evolve through a combination of classical evolution and quantum effects such as stochastic inflation and quantum tunnelling.
Some regions of the universe may become trapped in metastable minima with positive vacuum energy creating a large region of inflating false vacuum.
Since the field can subsequently tunnel out of this false vacuum region through nucleation of a bubble, open inflation is naturally embedded into the global dynamics.

In this paper, we restrict ourselves to collisions in either one- or two-field models.
The single-field approximation to the landscape is common in the existing literature, and we demonstrate explicitly that the transition from single field models through the addition of even a single field can dramatically change the outcomes of collisions.
Thus, we encourage some caution in drawing direct conclusions about bubble collisions on a very high-dimensional landscape.
Furthermore, as mentioned above the full dynamics on a complicated potential will consist of much more than just bubble nucleations and in some regimes quantum interference effects may be important.
We will not concern ourselves with these global issues and assume that in some region of the landscape the approximation of well-defined pairwise bubble collisions is adequate.

Many studies have considered the phenomenology of open inflationary models arising from bubble nucleation~\cite{Gott:1982zf,Bucher:1994gb,GarciaBellido:1997uh,Linde:1998iw,Freivogel:2005vv,Sugimura:2011tk}.
Collisions between vacuum bubbles have also been considered, beginning with the work of Hawking, Moss and Stewart~\cite{Hawking:1982ga}.
Early studies were motivated by the dynamics of early phase transitions within our horizon, with much of the focus on the production of gravitational waves~\cite{Turner:1992tz,Watkins:1991zt,Kosowsky:1991ua,Kosowsky:1992vn,Kosowsky:1992rz,Kamionkowski:1993fg,Child:2012qg,Kim:2014ara}.
Recent studies are more often motivated by bubble nucleation in false vacuum inflation~\cite{Aguirre:2008wy,Aguirre:2009ug,Johnson:2010bn,Freivogel:2007fx,Kleban:2011pg,Easther:2009ft}.
These investigations have culminated in a general relativistic treatment of the bubble problem~\cite{Johnson:2011wt,Hwang:2012pj,Wainwright:2013lea,Wainwright:2014pta}.
However, all of these studies assume the dynamics obeys an SO(2,1) symmetry, and no consideration is given to perturbations that break the symmetry.
Even studies of fluctuations around the SO(2,1) collision restrict to perturbations that can still be evolved using the approximation of 1+1-dimensional field theory~\cite{Gobbetti:2012yq,Czech:2010rg,Chang:2007eq}.
For the case of a single bubble the evolution of a more general set of fluctuations has been considered~\cite{Adams:1989su,Garriga:1991ts,Garriga:1991tb}.

In an effort to observationally constrain such scenarios, several recent studies have proposed observational signatures from the collision between our ``observation'' bubble and an external ``collision'' bubble~\cite{Aguirre:2007an,Aguirre:2007wm,Chang:2008gj,Kleban:2011yc,Zhang:2015uta}.
Several data searches have also been performed in conjunction with theoretical work, but thus far no signal has been found~\cite{Feeney:2010jj,Feeney:2010dd,McEwen:2012uk,McEwen:2012sv,Feeney:2012hj,Osborne:2013jea,Osborne:2013hea}.

Since observing a collision with another bubble universe would revolutionize our understanding of the cosmos, it is important to properly assess the possible outcomes of collisions.
As well, given the fundamental nature of bubble nucleations to any first-order phase transition, a thorough understanding of the collision dynamics is vital to a complete understanding of such transitions.
Of particular relevance to us is the validity of the SO(2,1) symmetry assumption, which has never been explicitly checked.
The results presented in this paper give a more complete understanding of bubble collisions in scalar field theories,
without relying on the assumption of SO(2,1) symmetry to simplify the problem.
We demonstrate that in a broad class of collisions, the dynamical amplification of quantum fluctuations leads to a complete breaking of the SO(2,1) symmetry.
Since the SO(2,1) assumption severely constrains the form of the final observable signals, it is possible that the searches performed thus far (which were based on studies assuming SO(2,1) symmetry) have been blind to other interesting possibilities.
Although we focus on the dynamics of the collisions rather than the final observable signatures, 
our results suggest several interesting new observational avenues for testing open inflation.

The remainder of the paper is organized as follows.
In section~\ref{sec:init_cond} we present our models and outline a pseudospectral approach to solve for the initial bubble profile.
Section~\ref{sec:one_bubble} considers the evolution of individual bubbles, both in the thin-wall and thick-wall regimes.
A brief review of the collision dynamics under the assumption of SO(2,1) symmetry is given in section~\ref{sec:symmetric_collision}.
We demonstrate that our lattice simulations reproduce the expected behaviour and test the sensitivity of the dynamics to the accuracy of the initial bubble profile.
Our main results are then given in section~\ref{sec:one_field} where we study collisions between bubbles in a variety of single-field models, including small initial fluctuations around the SO(2,1) profile.
We extend these findings to a simple two-field model that permits inflation inside the bubble in section~\ref{sec:two_field}.
Some brief comments on possible observable signatures and extensions to other scenarios are given in sections~\ref{sec:observations} and~\ref{sec:extensions}.
Finally we conclude in section~\ref{sec:conclusions_bub}.

\section{Single Field Models and Description of Initial Conditions}
\label{sec:init_cond}
To provide a concrete setting we focus primarily on double-well potentials with broken $Z_2$ symmetry.\footnote{Additional potentials are introduced later in the text as needed.}
We make two choices for the symmetry breaking
\begin{equation}
  V_{linear}(\phi) = \frac{\lambda}{4}\left( \phi^2 - \phi_0^2 \right)^2 - \delta \lambda\phi_0^3\left(\phi - \phi_0\right) + V_0
  \label{eqn:potential_linear}
\end{equation}
and
\begin{equation}
  V_{cubic}(\phi) = \frac{\lambda}{4}\left( \phi^2 - \phi_0^2 \right)^2 + \delta\lambda \phi_0^4 \left(\frac{\phi^3}{3\phi_0^3} - \frac{\phi}{\phi_0} + \frac{2}{3}\right) + V_0
  \label{eqn:potential_cubic}
\end{equation}
which we refer to as the linear and cubic symmetry breaking potentials respectively.
In both cases $\delta$ controls the difference between the false and true vacuum energies and $V_0$ can be adjusted to give the desired true vacuum energy.  
Our investigations here are restricted to Minkowski space, so $V_0$ does not influence the dynamics.\footnote{We also performed several runs with a homogeneous background Hubble and found no qualitative change to our results.}
In order for solutions describing vacuum tunnelling to exist in Minkowski spacetime, the false vacuum and true vacuum must be nondegenerate.
For the linear potential~\eqref{eqn:potential_linear}, the second minimum disappears for $|\delta| \geq 2/\sqrt{3}$, while for the cubic potential~\eqref{eqn:potential_cubic} this occurs when $|\delta| \geq 1$.
In the linear potential, the locations of the two minima depend on $\delta$, while they are fixed at $\pm \phi_0$ for the cubic potential.
Unless explicitly stated, dimensionful parameters are measured in units of $m\equiv \sqrt{\lambda}\phi_0$ with the exception of the field measured in units of $\phi_0$ and the energy density $\rho$ measured in units of $\lambda\phi_0^4$.
\begin{figure}
  \centering
  \includegraphics[width=0.48\linewidth]{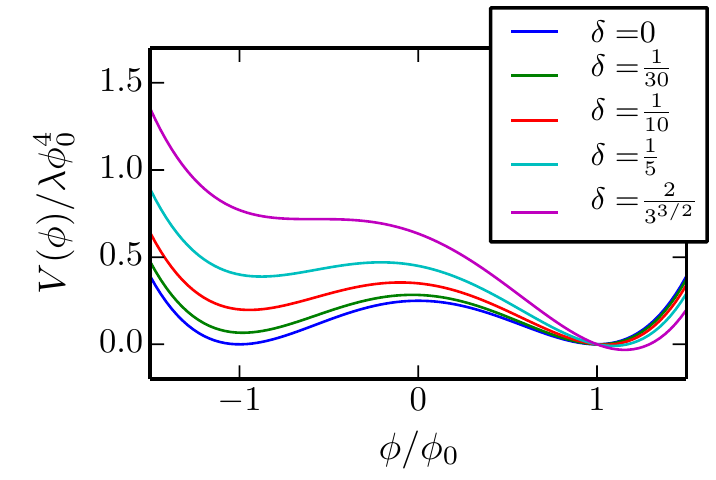}
  \includegraphics[width=0.48\linewidth]{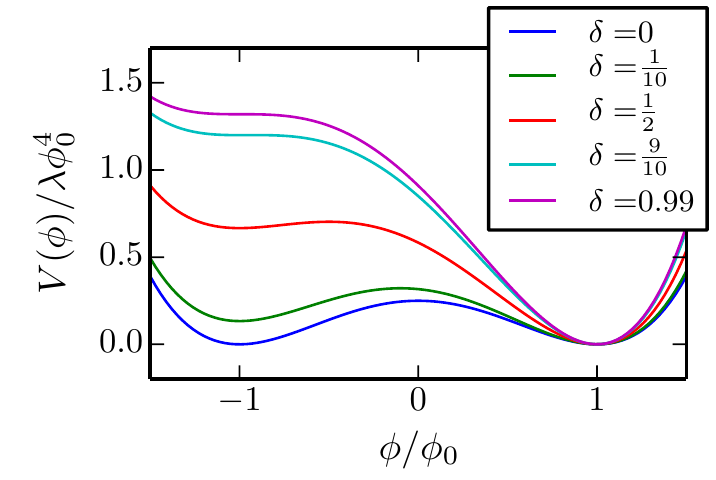}
  \caption[Plots of the potential for several choices of $\delta$]{Plots of the symmetry breaking potentials for several choices of $\delta$.  On the left we show the linear symmetry breaking~\eqref{eqn:potential_linear} and on the the right the cubic symmetry breaking~\eqref{eqn:potential_cubic}.  The largest value of $\delta$ in the linear potential is that for which the second minimum disappears.}
\end{figure}

\subsection{Solution of the Instanton Equation}
Before solving for the dynamics of bubble collisions, we must first find the initial conditions describing the profiles of the bubbles at nucleation.
In this subsection we present a new and extremely accurate numerical approach to determine the shape of the nucleated bubbles.
Throughout this paper we assume the \emph{most likely} bubble to nucleate within a surrounding false vacuum possesses an $SO(4)$ symmetry in Euclidean time and is described using the bounce formalism of Coleman~\cite{Coleman:1977py,Callan:1977pt,Coleman:1980aw}.
In Minkowski space, the profile satisfies the Euclidean signature equation
\begin{equation}
  \frac{\partial^2\phi}{\partial r_E^2} + \frac{3}{r_E}\frac{\partial\phi}{\partial r_E} - \frac{\partial V}{\partial\phi} = 0
  \label{eqn:instanton}
\end{equation}
with the boundary conditions
\begin{equation}
  \phi(r_E=\infty) = \phi_{\mathrm{false}} \qquad \frac{\partial\phi}{\partial r_E}(r_E=0) = 0 \, .
\end{equation}
Here $r_E^2 = d\tau^2 + d{\bf x}^2$ is the Euclidean radius and $\tau=it$ is the Euclidean time.
For a single scalar field in Minkowski space with potential satisfying some regularity conditions, it has been shown that the solution to~\eqref{eqn:instanton} is the minimum action solution relevant to false vacuum decay~\cite{Coleman:1977th}.

Before presenting our technique, we briefly review the most common analytic approach (the thin-wall approximation) and numerical approach (the overshoot-undershoot method) used to solve~\eqref{eqn:instanton}.
In the thin-wall limit, valid if the initial radius of the bubble is much greater than the width of the wall, we can obtain simple analytic estimates for the bounce profile.
For the potentials~\eqref{eqn:potential_linear} and~\eqref{eqn:potential_cubic}, this approximation is valid when $\delta \ll 1$.
In the simplest form of the approximation, we drop the friction term and the terms proportional to $\delta$ in the equation of motion. 
This leaves the equation for a domain wall in the degenerate double well $\frac{\lambda}{4}\left(\phi^2-\phi_0^2\right)^2$.
Performing the usual quadrature gives an expression for the field profile
\begin{equation}
  \phi_{wall}=\phi_0\tanh\left(\frac{r-R_0}{\sqrt{2}}\right) \, .
\end{equation}
We then adjust the multiplier on the $\tanh$ and add a constant so that the field interpolates between the false and true vacuum.
Finally, conservation of energy gives the initial bubble radius as 
\begin{equation}
  R_0 = \frac{3\sigma}{\Delta\rho} \qquad \mathrm{where} \qquad \sigma = \int dr [\partial_r\phi_{wall}(t=0)]^2
\end{equation}
is the surface tension of the bubble wall and $\Delta \rho = V(\phi_{false})-V(\phi_{true})$ is the difference in potential energy between the false and true vacuum.
For the linear potential~\eqref{eqn:potential_linear} we have 
\begin{equation}
  R_{0}^{linear} = \sqrt{2}\delta^{-1}
\end{equation}
and for the cubic potential~\eqref{eqn:potential_cubic} we have 
\begin{equation}
  R_{0}^{cubic}=\frac{3}{\sqrt{2}}\delta^{-1} \, .
\end{equation}
It should be clear that with some modifications the thin-wall approximation can be applied to other potentials.
In particular, if we replace $\phi_{true}$ with the location where the field tunnels out ($\phi_{tunnel}$), we can apply the thin-wall approximation in situations where the field does not tunnel out near the true vacuum.
Of course, we must either estimate $\phi_{tunnel}$ or else determine it by solving the bounce equation~\eqref{eqn:instanton} in order to obtain the initial radius.
If the latter approach is used to find $\phi_{tunnel}$, then the thin-wall approximation itself will provide little utility other than aiding intuition.

Although the simplicity of the final result makes the thin-wall approximation attractive, 
there are situations where it is either invalid or else insufficiently accurate.
We must then turn to a numerical solution of~\eqref{eqn:instanton}.
In the literature, the equation is usually solved via a shooting method.
The ODE is recast into a root finding problem $\vec{\phi}(r_E=\infty|\vec{\phi}(r_E=0)=\vec{\phi}_0) - \vec{\phi}_{false} = 0$ for the initial field value $\vec{\phi}_{0}$ at $r_E = 0$.
For the single-field case, the root finding is often done via bisection and the resulting algorithm is known as the overshoot-undershoot method.
To extend this to the multifield case one needs to use a root-finder that works in multiple dimensions.
An obvious choice is a Newton-type method,
with the required derivatives obtained by solving for a collection of nearby trajectories and then doing a polynomial interpolation.
Of course, the desired solution is a saddle point and care must be taken to evaluate these derivatives before the neighbouring trajectories diverge away from the target solution.
One possible way to extend the overshoot-undershoot approach to multiple fields can be found in~\cite{Wainwright:2011kj}.

\begin{figure}[t]
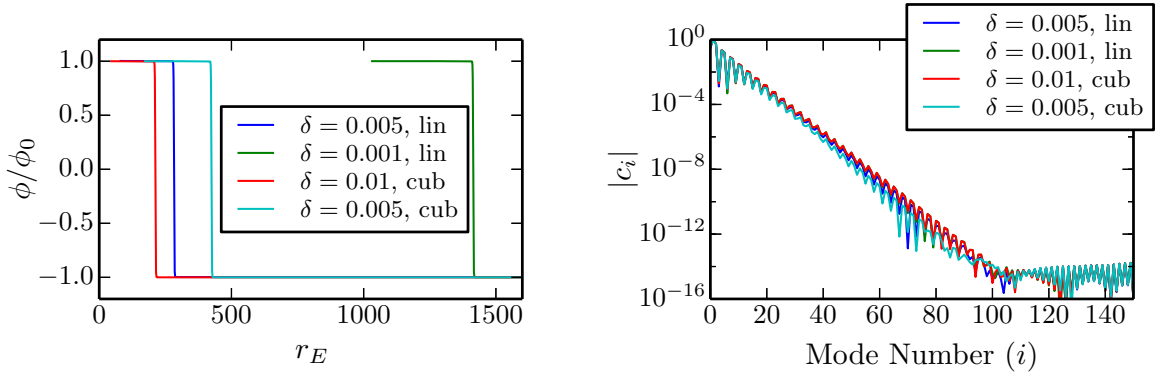

  \includegraphics[width=0.48\linewidth]{{{thin_wall_bubbles}}}
  \hfill
  \includegraphics[width=0.48\linewidth]{{{thin_wall_coeffs}}}
  \caption[Instanton solutions for some extremely thin-walled bubbles.]{Numerical solutions to~\eqref{eqn:instanton} in the thin-wall limit.  For each case the ratio of the wall width ($mw = \sqrt{2}$) to the initial bubble radius is $w/R_{0} \sim \delta$.  In the left panel we show the instanton profiles (zoomed into the region $r_{E} \lesssim R_{0}$), while in the right panel we show the resulting spectral coefficients.  For all cases, we took $L=\sqrt{3}R_{0}$ with $R_{0}=\sqrt{2}\delta^{-1}$ for the linear potential and $R_{0} = 1.5\sqrt{2}\delta^{-1}$ for the cubic potential.  Meanwhile, the ``stretching'' parameter $d$ was $d=0.022, 0.0045$ for $\delta=0.005,0.001$ in the linear potential, and $d=0.03, 0.014$ for $\delta=0.01,0.005$ in the cubic potential.}
  \label{fig:step_thinwall_instantons}
\end{figure}

\begin{figure}
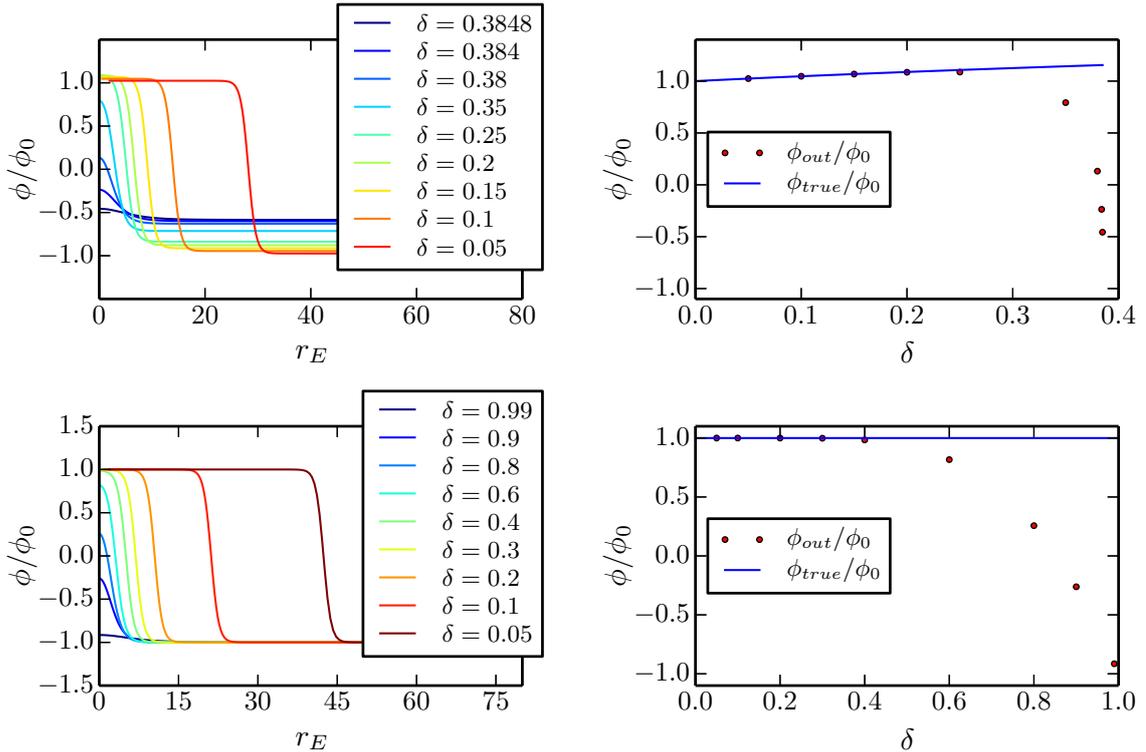

  \centering
  \begin{tabular}{cc}
  \includegraphics[width=0.48\linewidth]{{{instanton_1field_linear_varydelta}}} &
  \includegraphics[width=0.48\linewidth]{{{phitunnel_linear_varydelta}}} \\

  \includegraphics[width=0.48\linewidth]{{{instanton_1field_cubic_varydelta}}} &
  \includegraphics[width=0.48\linewidth]{{{phitunnel_cubic_varydelta}}}
  \end{tabular}
  \caption[Instanton profiles for a wide range of symmetry breaking parameters $\delta$ and deviation of the tunnel out location of the field from the minimum]{In the top row we show instanton profiles in the linearly asymmetric double well~\eqref{eqn:potential_linear} for a range of $\delta$ values.
The left panel shows the instanton profiles (zoomed into the region where the field is varying).
Meanwhile the right panel shows $\phi(r_E=0)$ which becomes the initial field value at the center of the bubble.  For comparison we also include the location of the true vacuum minimum.
The bottom row shows the same two plots for the cubically broken symmetry~\eqref{eqn:potential_cubic}.
For $\delta \gtrsim 0.25, 0.4$ the location where the field tunnels out to begins to deviate significantly from the false vacuum.}
  \label{fig:instantons_varydelta}
\end{figure}

Rather than adopt a shooting method, we instead use a pseudospectral approach and expand the function in even rational Chebyshev functions on the interval $(-\infty,\infty)$
\begin{align}
  \notag \phi_{bounce}(r_{E}) = \sum_i c_i B_{2i}\left(y\left(\frac{r_E}{\sqrt{r_E^2+L^2}}\right) \right) \\
  y(x) = \frac{1}{\pi}\tan^{-1}\left( d^{-1}\tan\left(\pi\left[x-\frac{1}{2}\right]\right) \right) + \frac{1}{2}
  \label{eqn:spectral_expansion}
\end{align}
where $B_n$ are the Chebyshev polynomials.
Since this is a global expansion method, it displays exponential convergence as the number of lattice sites is increased and machine-precision accuracy is easily obtained.
Without the mapping $y(x)$ (\ie\ setting $y(x)=x$), this is simply an expansion in the rational Chebyshev functions $TB_n(y) = B_n\left(\frac{y}{\sqrt{y^2+L^2}}\right) = \cos\left(n\cot^{-1}\left(\frac{y}{L}\right)\right)$ on the doubly-infinite interval.  
$L$ is an adjustable parameter that determines where the oscillations in the rational Chebyshev functions (or equivalently the collocation points) are clustered on the infinite interval.
The even $TB(x)$'s form a complete set for smooth functions which asymptote to a constant at infinity 
and are symmetric about the origin (such as the bounce).
Thus they enforce the $r_E=0$ boundary condition automatically.
We also include an additional mapping $y(x)$ which clusters (repels) points around $r_E=\frac{L}{\sqrt{3}}$ if $d<1$ ($d>1$) without simultaneously introducing new singularities at the boundaries which would greatly reduce the convergence rate of the expansion.
Via a judicious choice of $L$ and $d$, we can resolve instantons to machine precision even in the extremely thin-wall case.
We illustrate the efficiency of this expansion in~\figref{fig:step_thinwall_instantons}, where we show instantons for four extremely thin-walled bubbles as well as the spectral coefficients $c_i$ in~\eqref{eqn:spectral_expansion}.
Numerically this is the most difficult case if no finesse is used in the solution.
In all cases the round-off plateau for the coefficients is clearly present at $i \approx 100$, indicating we have hit the limits of double-precision arithmetic.
Through the use of a smart collocation grid, we manage to achieve this accuracy using fewer modes than the ratio of the bubble radius to its width $\frac{R_{init}}{w}$.
As an added bonus, our outer collocation point is located at $r_E=\infty$ so there are no errors introduced by putting the system in a finite box.
To demonstrate the utility of this approach across a range of potential deformations,
~\figref{fig:instantons_varydelta} shows the instanton profiles in the linear~\eqref{eqn:potential_linear} and cubic potential~\eqref{eqn:potential_cubic} for a range of $\delta$.

Of course, the bounce solution only describes the most likely initial profile for the field.
There are additional fluctuations which we include by
taking the initial conditions for a single bubble to be
\begin{equation}
  \phi_{init}({\bf x},0) = \phi_{bounce}(r_E=|{\bf x}|) + \delta\phi \qquad \dot{\phi}_{init}({\bf x},0) = \delta\dot{\phi}
\end{equation}
where $\phi_{bounce}$ is a solution of~\eqref{eqn:instanton} and $\delta\phi$ and $\delta\dot{\phi}$ are realizations of random fields.
This is generalized to multibubble initial conditions in the obvious way.
The fluctuations $\delta\phi$ encapsulate the effects of quantum fluctuations: deviations of the nucleated bubble wall from perfect SO(3,1) symmetry, bulk fluctuations in the ambient false vacuum, and bulk fluctuations in the bubble interior.
Since we consider scenarios where the bubbles form via quantum nucleation, it is inconsistent to ignore the fluctuations since rare coherent excursions of these fluctuations are what allows nucleation to occur at all. 

A proper determination of the initial conditions requires a calculation of how the fluctuations in the original false vacuum are projected into the bubble spacetime by the nucleation event (see e.g.~\cite{Garriga:1991ts,Garriga:1991tb,Sasaki:1994yt}).
This calculation is beyond the scope of this paper and would only serve to obscure the essence of our result.
Instead we simply initialize the fluctuations as a realization of a homogeneous Gaussian random field with spectrum
\begin{equation}
  \langle|\delta\tilde{\phi}_k|^2\rangle \sim \frac{1}{2\sqrt{k^2+V''(\phi_{fv})}} \qquad   \langle|\delta\tilde{\phi}_k|^2\rangle \sim \frac{\sqrt{k^2+V''(\phi_{fv})}}{2} \, 
\end{equation}
to mimic the fluctuations if the field were sitting at its false vacuum mimimum in Minkowski space.\footnote{As a check of this assumption, we also ran simulations with several other choices of initial fluctuations.  One set included homogeneous bulk fluctuations with different initial spectra than the Minkowski vacuum.  Another set involved initializing $\phi_{\mathrm{init}} = \phi_{\mathrm{bounce}}(r_E=|{\bf x}|+\delta r)$ with $\delta r =\sum_{\ell m} a_{\ell m}Y_{\ell m}$ a 2d random field obtained by realizing a collection of $a_{\ell m}$'s.  Again, we used several different spectra for our $a_{\ell m}$'s to verify that our results were not sensitive to a particular choice.  In every case we tried, the outcome of a collision of two bubbles was qualitatively the same as the results we present in the remainder of the paper.}
The overall scale of the potential $\lambda$ enters into the initial fluctuation amplitude as $\frac{\delta\phi}{\phi_0} \propto \sqrt{\lambda}$.
We measure the fields in units of $\phi_0$ and time in units of $\left(\sqrt{\lambda}\phi_0\right)^{-1}$, so the vev $\phi_0$ only appears in the equations of motion if we consider coupling to gravity where it determines the strength of the gravitational interaction via $\frac{\phi_0}{M_P}$.
We use the same convolution based method as DEFROST to initialize the fluctuations~\cite{Frolov:2008hy,Pen:1997up}.

The remainder of the paper presents results obtained from a massively parallel scalar field lattice code written by one of the authors.
Hamilton's equations for the discretized system were evolved using a sixth-order symplectic Yoshida integrator~\cite{Yoshida:1990,Huang:2011gf,Sainio:2012mw}
and (unless otherwise indicated) a second-order accurate and fourth-order isotropic stencil for the \laplacian~\cite{Frolov:2008hy,Patra:2006}.
All production runs used lattices with $1024^3$ sites, although we did perform some numerical checks using $2048^3$ lattices.
For all cases, the total energy of the system (when running Minkowski simulations) was conserved to the $10^{-9}$ level or better.

To provide a test of our lattice simulations and facilitate comparison with previous studies,
 we also present the results of dimensionally reduced 1+1-dimensional simulations at several points in the subsequent sections.
These simulations used a Fourier pseudospectral lattice discretization and a 10th order Gauss-Legendre time integrator.
The basics of this approach are outlined in appendix B of~\cite{ref:bbm1}.

\section{Evolution of a Single Bubble}
\label{sec:one_bubble}
\begin{figure}
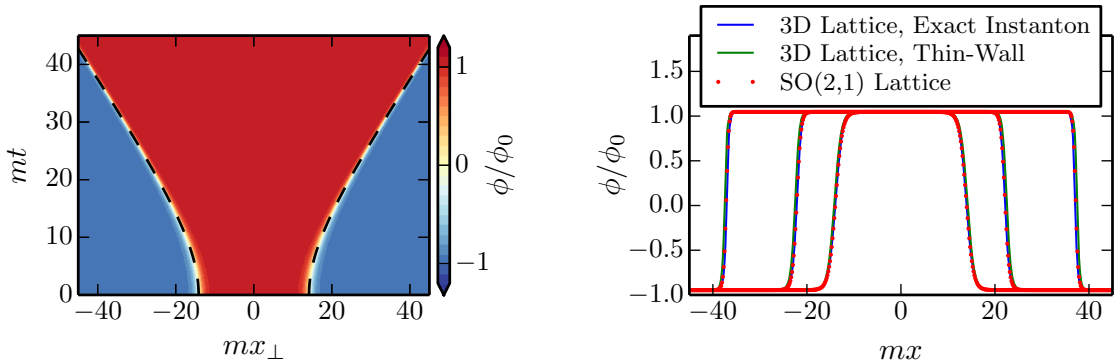

  \centering
  \includegraphics[width=0.48\linewidth]{{{one_bubble_linear_del0.1_plane}}}
  \hfill
  \includegraphics[width=0.48\linewidth]{{{one_bubble_linear_del0.1_tslices}}}
  \caption[Time evolution of a thin-walled bubble with $\delta=0.1$ in the linear symmetry breaking potential]{Time evolution of a thin-walled bubble with $\delta=0.1$ in the linear symmetry breaking potential. In the left panel we show the field value as a function of $mt$ and position $mx_\perp$ along a slice taken through the center of the bubble.  This figure is from a three-dimensional lattice simulation.  For reference, the location of the wall in the thin-wall approximation is included as a dashed black line.  In the right panel, we show field profiles at time $mt=0,17.24,34.48$ on a slice through the center of the bubble.  To demonstrate the accuracy of our numerics, we compare our lattice results to higher-resolution one-dimensional simulations.  For comparison, we also include the evolution of a bubble whose initial profile is given by the thin-wall approximation rather than an exact numerical calculation.}
  \label{fig:thin_wall_singlebubble}
\end{figure}

First we consider the evolution of individual bubbles.
Fluctuations in the angular dependence of the bubble radius for the case of individual bubbles in the thin-wall limit were studied and found to be stable in~\cite{Adams:1989su,Garriga:1991tb}.
Our results in this section demonstrate that this is also true for the bulk fluctuations around \emph{single} bubbles.
We evolve a thin-walled bubble with $\delta=0.1$ in the linear potential and a thick-walled bubble with $\delta=0.99$ in the cubic potential.
The corresponding instanton profiles can be found in the left panels of~\figref{fig:instantons_varydelta}.
For the thin-walled bubble, the field tunnels out very close to the true vacuum.
As a result, we can think of this as a bubble with true vacuum interior separated from the false vacuum exterior by a domain wall of width $m^{-1}$.
For the single bubble this wall can be approximated as infinitely thin.
The pressure differential between the interior and exterior of the bubble then causes the wall to accelerate and it follows a hyperbolic curve given by
$r_{wall}(t) = \sqrt{r_{wall}(t=0)^2 + t^2}$ as shown in~\figref{fig:thin_wall_singlebubble}.

\begin{figure}
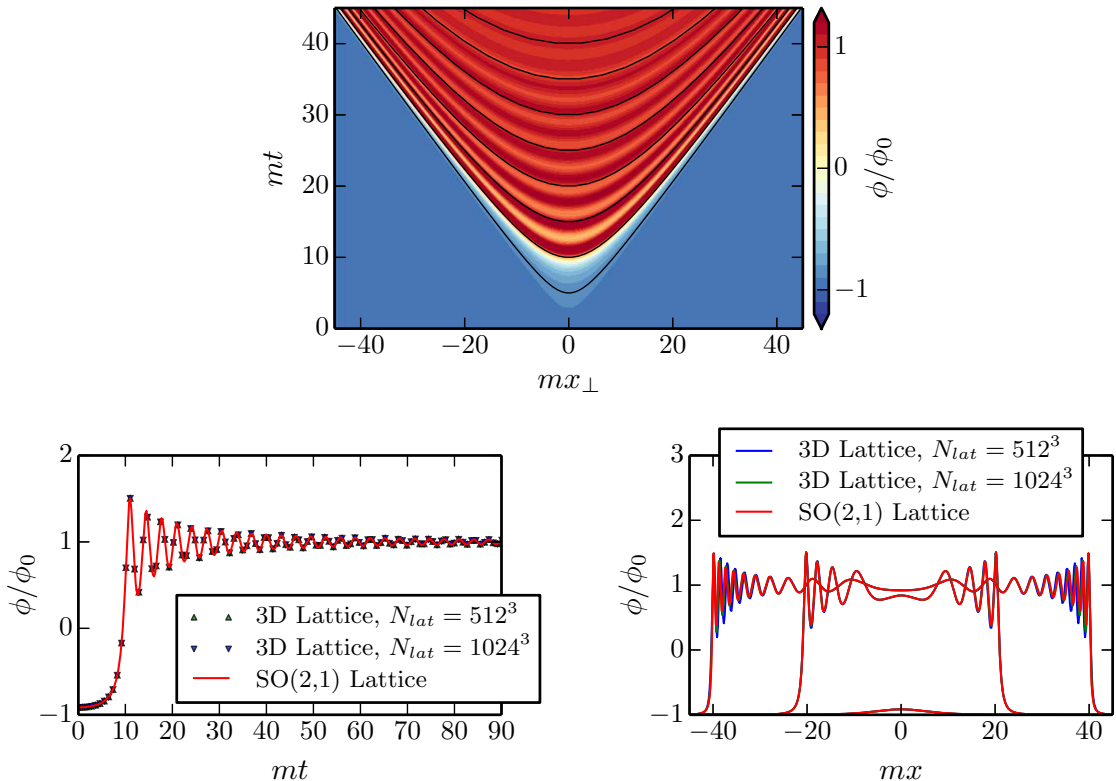

  \begin{center}
  \includegraphics[width=0.6\linewidth]{{{one_bubble_del0.99_transverse}}}\\
  \includegraphics[width=0.48\linewidth]{{{one_bubble_del0.99_center}}}
  \hfill
  \includegraphics[width=0.48\linewidth]{{{one_bubble_del0.99_latet}}}
  \end{center}
  \caption[Time evolution of a single thick-walled bubble with $\delta=0.99$ in the cubic symmetry breaking potential]{Time evolution of a single thick-walled bubble with $\delta=0.99$ in the cubic symmetry breaking potential.  In the top panel, we plot the field as a function of $mt$ and position $mx_\perp$ along a line through the center of the bubble.  Regions where the field is near the false vacuum are blue, while regions where the field is near the true vacuum are red.  In the bottom left panel, we plot the value of the field at the center of the bubble as a function of time.  As expected the field undergoes damped harmonic oscillations.  As a test of our numerical code, we include the result for two different choices of lattice spacing $dx$, and also for a one-dimensional simulation using a pseudospectral discretization and 10th order Gauss-Legendre time integrator.  In the bottom right panel we show the field profile at times $mt=0,22.98$ and $41.37$, again for the two different lattice grid spacings and also the one-dimensional simulation.}
  \label{fig:thick_wall_onebubble}
\end{figure}

\begin{figure}
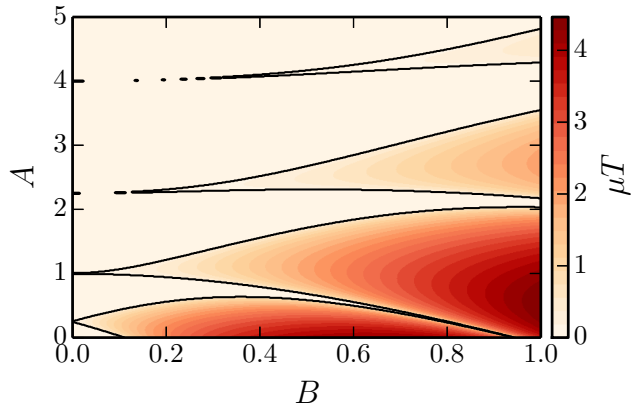

  \begin{center}
  \includegraphics[width=0.6\linewidth]{{{floquet_thick_wall}}}
  \end{center}
  \caption[Floquet chart for oscillations in the interior of a thick-walled bubble]{Floquet chart for $\ddot{f} + \left(A + 4B\sin(t) + 1.5B^2\sin^2(t)\right)f=0$, corresponding to oscillations about the true vacuum minimum for $\delta=1$.  For oscillations in the interior of the thick-walled bubble, we have $A \geq 2$ and $B \lesssim 0.5$.}
  \label{fig:thick_wall_floquet}
\end{figure}

For the thick-walled bubble the evolution is somewhat different as seen in~\figref{fig:thick_wall_onebubble}.
The field now tunnels out far from the true vacuum.
In the bubble interior it begins to oscillate around the minimum at $\phi=\phi_0$.
In the standard slicing of Minkowski space used in the code, these oscillations appear as outward propagating spherical waves.
The leading edge of this spherical wave quickly develops into the bubble wall and propagates outward with a speed asymptotically approaching the speed of light.
This is easiest to see by foliating the spacetime with hyperboloids centered on the nucleation site of the bubble.
We label each of these hyperboloids by a new time coordinate $\tilde{t}$ related to the time coordinate used in our simulations via $t=\tilde{t}\cosh\chi$ 
where $\chi$ is the radial coordinate along the hyperboloids.
The line element in the new slicing is $ds^2=-d\tilde{t}^2+\tilde{t}^2dH_3$ with $H_3$ the unit three-hyperboloid.
For the exact instanton initial condition, the evolved field is a function of $\tilde{t}$ only and satisfies
\begin{equation}
  \frac{\partial^2\phi}{\partial \tilde{t}^2} + \frac{3}{\tilde{t}}\frac{\partial\phi}{\partial \tilde{t}} + V'(\phi) = 0 \, .
\end{equation}
In this particular case the oscillating field in the interior of the bubble does not lead to a strong preheating instability because the potential seen by the field as it oscillates in the interior of the bubble is very nearly quadratic.\footnote{For different choices of potential or couplings to additional fields the interior of the bubble may experience strong instabilities (see for e.g.~\cite{Felder:2001kt}).}
To see this explicitly, define $\psi = \phi - \phi_0$ and reexpand the potential to obtain
\begin{equation}
  U(\psi) \equiv V(\phi_0+\psi)= (\lambda+\delta)\phi_0^2\psi^2 + \frac{(3\lambda+\delta)}{3}\phi_0\psi^3 + \frac{\lambda}{4}\psi^4 \, .
\end{equation}
The equation for linear perturbations around $\psi(\tilde{t})$ is then
\begin{equation}
  \frac{\partial^2(\tilde{t}^{3/2}\delta\psi)}{\partial (m_{eff}\tilde{t})^2} + \left(\frac{\kappa^2}{m_{eff}^2\tilde{t}^2} + (1+\delta) + (3+\delta)\frac{\psi}{\phi_0} + \frac{3}{2}\frac{\psi^2}{\phi_0^2} - \frac{3}{4\tilde{t}^2}\right)(\tilde{t}^{3/2}\delta\psi) = 0 
\end{equation}
where $m_{eff}^2 = \partial_{\psi\psi}U(\psi=0) = 2\lambda\phi_0^2$ and $\kappa^2$ is an eigenvalue of the Laplacian on the unit three-hyperboloid.
To gain some intuition about these instabilities, let's approximate the motion of $\psi$ as it oscillates around the minimum by $\psi = \alpha\phi_0t^{-3/2}\sin(m_{eff}\tilde{t}+\theta_0)$ with $\alpha < 1$ a numerical coefficient and $\theta_0$ a phase.
At the center of the bubble the time coordinate $\tilde{t}$ coincides with the time parameter $t$ used in our simulations.
Therefore, we see from the middle panel of~\figref{fig:thick_wall_onebubble} that this parameterization of $\psi$ is indeed a good approximation.
Ignoring the time dependence of all coefficients, the fluctuations obey an equation of the form
\begin{equation}
  \partial_{\bar{t}\bar{t}}f + \left(A  + (3+\delta)B\sin(\bar{t}) + \frac{3}{2}B^2\sin^2(\bar{t}) \right)f=0 \, .
\end{equation}
For the continuum part of the spectrum, we have $\kappa^2 \geq 1$ 
so $A \geq (1+\delta)$ and from~\figref{fig:thick_wall_onebubble} we see that $B \lesssim 0.5$ during the oscillations.
In~\figref{fig:thick_wall_floquet} we see that this puts us well into the weak resonance regime.
Since the oscillations damp with time, a given mode will trace a line in the instability chart so that even in the long-time limit there is no exponential growth and the decay is perturbative~\cite{Braden:2010wd}.

\section{3D Bubble Collisions Without Bulk Fluctuations}
\label{sec:symmetric_collision}
In this section we study collisions between pairs of nucleated vacuum bubbles working under the assumption of SO(2,1) symmetry.
This allows us to compare our results with previous treatments of this problem (where the SO(2,1) symmetry is assumed to hold exactly and the dynamics are reduced to one spatial dimension).
These simulations also provide a highly nontrivial test of our numerical approach.
We consider both the case of the spectrally accurate numerical profiles for the bubbles, as well as initial profiles given by the thin-wall approximation.
Since the thin-wall profile is not an exact solution to the bounce equation~\eqref{eqn:instanton}, the latter choice is equivalent to a small breaking of the boost symmetry for the single bubbles.
As we demonstrate below, even this minimal breaking of the symmetries of the bubble can have a dramatic effect on the full three-dimensional evolution.
The nonlinear field dynamics results in a complete breakdown of the boost symmetries shortly after the collision.
This breaking of the boost symmetries is our first new result requiring three-dimensional (more precisely higher than one-dimensional) simulations.

\subsection{Evolution of SO(2,1) Background and Linear Evolution of Fluctuations}
\label{sec:bubbles_linear}
\begin{figure}[h]
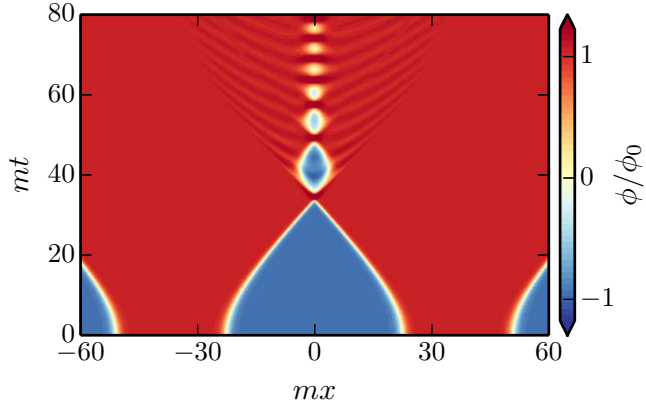

  \centering
  \includegraphics[width=0.6\linewidth]{{{bubble_1d_del0.1}}}
  \caption[Two bubble collision]{SO(2,1) symmetric collision between two bubbles in the linear symmetry breaking potential with $\delta=0.1$.  Red corresponds to regions where the field is near the true vacuum minimum and blue to regions where it is near the false vacuum minimum.  The two bubble walls do not immediately annihilate each other in the collision, but instead they undergo multiple bounces while slowly radiating energy into the bulk.}
  \label{fig:bubble_collision_symmetric}
\end{figure}

Before attacking the full problem, let's review linear stability analysis around the SO(2,1) symmetric solution~\cite{ref:bbm1}.
For the case of perturbations to a single bubble see~\cite{Adams:1989su,Garriga:1991ts,Garriga:1991tb}.
To make the symmetry manifest, define new coordinates
\begin{align}
  t &= s\cosh\chi \notag \\
  x &= x \notag \\
  y &= s\sinh\chi\cos\theta \notag \\
  \label{eqn:so2_coordinates}
  z &= s\sinh\chi\sin\theta \, .
\end{align}
Field configurations that depend only on $s$ and $x$ are SO(2,1) symmetric.
Now consider fluctuations around such a symmetric background
\begin{equation}
  \phi = \phi_{bg}(s,x) + \delta\phi(s,x,\chi,\phi) \, .
\end{equation}
In the limit of linear fluctuations, the background solution satisfies
\begin{equation}
  \frac{\partial^2\phi_{bg}}{\partial s^2} + \frac{2}{s}\frac{\partial\phi_{bg}}{\partial s} - \frac{\partial^2\phi_{bg}}{\partial x^2} + V'(\phi_{bg}) = 0
  \label{eqn:bubble_bg}
\end{equation}
and the fluctuations obey a one dimensional wave-equation with time- and space-dependent effective mass
\begin{equation}
  \frac{\partial^2 A_{\ell}}{\partial s^2} + \frac{2}{s}\frac{\partial A_{\ell}}{\partial s} - \frac{\partial^2 A_{\ell}}{\partial x^2} + 
  \left(\frac{\ell^2}{s^2} +  V''(\phi_{bg}) \right)A_{\ell} = 0 \, .
  \label{eqn:bubble_fluc}
\end{equation}
We expanded the perturbation in eigenmodes 
\begin{equation}
  \delta\phi = \sum_{\ell,n} A_{\ell}(s,x)C_{\ell,n}(\chi)e^{in\theta}\ \mathrm{with}\ n \in \mathbb{Z} \, .
\end{equation}
The $C_{\ell,n}$ labelled by $\ell$ are the eigenfunctions and eigenvalues of
\begin{equation}
    \frac{1}{\sinh(\chi)}\frac{d}{d\chi}\left(\sinh(\chi)\frac{dC}{d\chi}\right) = \left(-\ell^2 + \frac{n^2}{\sinh^2(\chi)}\right)C \, .
\end{equation}
Past studies based on exact SO(2,1) symmetry restrict the treatment to a study of~\eqref{eqn:bubble_bg}.
A sample collision between two bubbles in the linear symmetry breaking potential with $\delta=0.1$ and SO(2,1) symmetry imposed is shown in~\figref{fig:bubble_collision_symmetric}.
The bubble walls do not immediately annihilate each other upon colliding, but instead bounce off of each other multiple times. 
Each bounce opens up a pocket at the collision site where the field is localized near the false vacuum minimum.
The bouncing behaviour was first noted by Hawking, Moss and Stewart~\cite{Hawking:1982ga} and is typical of thin-wall bubble collisions in double-well potentials.
The bouncing behaviour can persist when gravitational interactions are included~\cite{Johnson:2011wt}. 
Due to the SO(2,1) symmetry, each pocket corresponds to an expanding torus in the full 3-dimensional evolution, where the field inside the torus is near the the false vacuum minimum.
In~\cite{ref:bbm2} we showed that long tubes with the field near the false vacuum in the interior (formed from the breakup of colliding domain walls) are unstable to collapse.
Hence, we expect that when the cylindrical symmetry is not perfect the torii will fracture into oscillons.
In addition, small amplitude fluctuations localized near the axis connecting the bubble centers will see an oscillating background field as the two bubble walls bounce off each other.
As a result, the fluctuations in these regions will be resonantly amplified, analogous to the situation with planar domain walls.
We will see in section~\ref{sec:one_field} that both of these behaviours manifest themselves in the fully 3+1-dimensional problem.

\subsection{3D Simulation of SO(2,1) Bubble Collsions}
Before proceeding to the case with bulk fluctuations around the pair of nucleated bubbles, we first present results using initial conditions that preserve the SO(2,1) symmetry up to errors induced from using a superposition of single instanton solutions.
In the absence of numerical errors, the resulting time-evolution preserves the full SO(2,1) symmetry.
Therefore, this is a highly nontrivial test of our numerics and also provides a nice visualization of the three-dimensional field profile.

\begin{figure}
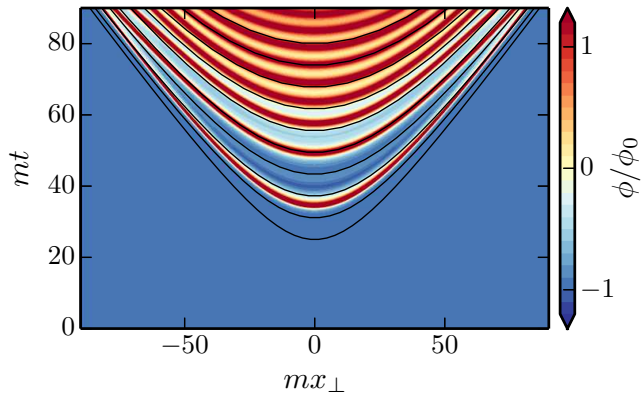

  \centering
  \includegraphics[width=0.6\linewidth]{{{bubble_exact_transverse_del0.1}}}
  \caption[Evolution of $\phi$ along the line $x=0=y$]{The evolution of $\phi$ along the line $x=0=y$ where we have chosen our coordinates such that the two bubbles first collide at the origin and the collision occurs along the $x$-axis.  For reference, we also include several lines of constant $s$, demonstrating that the boost symmetry is quite well preserved in the full three-dimensional simulation.}
  \label{fig:transverse_slice_nofluc}
\end{figure}

\begin{figure}
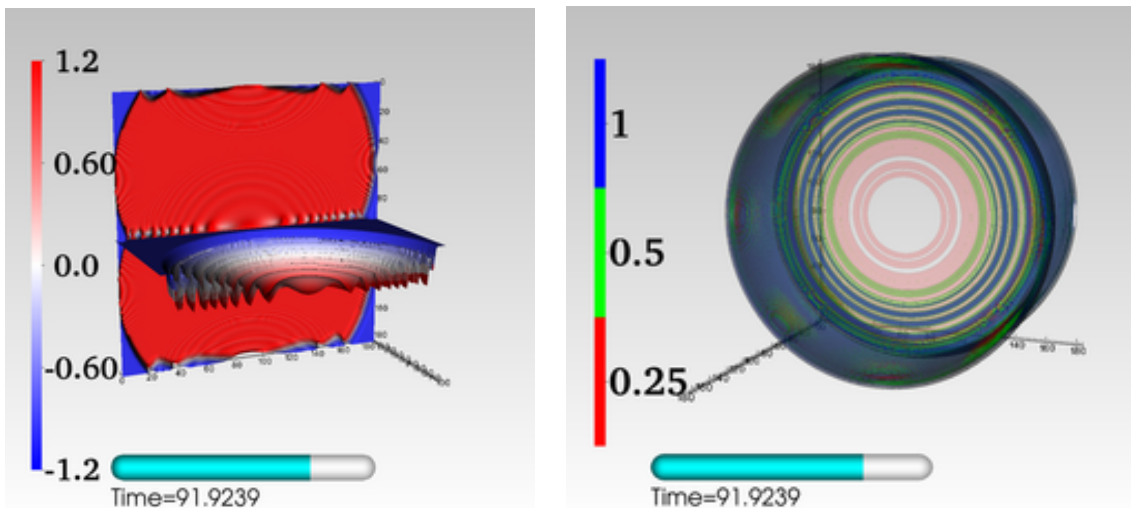

  \centering
  \begin{tabular}{cc}
    \includegraphics[width=0.45\linewidth]{{{bubble_field_twowell_exactic_0003}}} &
    \includegraphics[width=0.48\linewidth]{{{bubble_contour_twowell_exactic_0003}}}
  \end{tabular}
  \caption[Slice of the field and contours of the energy density $\rho/\lambda\phi_0^4$ for exact instanton initial condition.]{The post-collision field configuration for a two bubble collision in the linear symmetry breaking potential with $\delta=0.1$ and exact initial bubble profiles.  \emph{Left}: Two slices of the field at $mt=92$.  The vertical slice is taken along the collision axis and cuts through the centers of the bubbles, while the horizontal slice is taken orthogonal to the collision axis at the collision site. \emph{Right}: Contours of the energy density $\frac{\rho}{\lambda\phi_0^4}$, showing the expanding concentric torii whose interiors are near the false vacuum and whose walls interpolate between the false vacuum interior and the true vacuum exterior.  An animation for the evolving field profile can be found at \thinwallexactfield.}
  \label{fig:field_and_rho_nofluc}
\end{figure}

We consider our fiducial thin-walled instanton with $\delta=0.1$ in the linear symmetry breaking potential~\eqref{eqn:potential_linear}.
The single instanton profile can be seen in the top left panel of~\figref{fig:instantons_varydelta}.
From~\figref{fig:transverse_slice_nofluc} we see that the lattice preserves both the rotational and boost symmetries quite well, especially considering that spatial discretization and discrete time steps explicitly break both symmetries.
For times $mt \gtrsim 80$ the surfaces of constant field are deformed slightly from being perfect hyperboloids near the edges of the collision, but they still exhibit the correct qualitative form.
A late time three-dimensional view of the same collision as~\figref{fig:bubble_collision_symmetric} can be found in~\figref{fig:field_and_rho_nofluc}, where the preservation of the rotational symmetry about the collision axis is clear.
The expanding torii formed from the excursions of the field back to the false vacuum in the collision region are also clearly visible.
The total energy of the system is conserved to the $10^{-9}$ level, and the deviation from perfect boost symmetry persists when we halve the time step or change the order of the time integrator.
This indicates that the mild breaking of the boost symmetry is an artifact of the modified dispersion relationship of the finite-difference discretization rather than the time stepping.

\subsection{Sensitivity of Boost Symmetry to Initial Conditions}
\label{sec:boost_breaking}
We now consider the sensitivity of our results to initial perturbations by breaking the boost symmetries while still preserving the rotational symmetry about the collision axis.
Specifically, we perturb the instanton solution used to set initial conditions while assuming it is still only a function of $r_E$.  
The resulting instanton profile no longer satisfies~\eqref{eqn:instanton}, and therefore the resulting evolution of a single bubble is not boost invariant.
However,  the three rotational symmetries are preserved.
To see that the boost symmetries are broken, suppose that we choose a single bubble initial field configuration $\phi_{init,sym}({\bf x},0)$ such that the time-evolved field is a function of $|{\bf x}|^2-t^2$ alone (\ie\ it is SO(3,1) invariant).
Upon Wick rotation into Euclidean time, the SO(3,1) invariance translates into an SO(4) invariance so that the resulting Euclidean profile is a function of $r_E$ alone.
As well, since the time evolved profile is obtained by solving the Klein-Gordon equation in Lorentzian signature, the Euclidean profile must satisfy the Euclidean version of the Klein-Gordon equation.
However, for a function of $r_E$ alone the Euclidean Klein-Gordon equation is simply the equation for the bounce~\eqref{eqn:instanton}.\footnote{This argument does not rely on the boundary conditions for the bounce solutions being met, only that it satisfies the correct equation.} 
Since our perturbed initial condition does not satisfy~\eqref{eqn:instanton}, the resulting evolved bubble profile cannot respect SO(3,1) symmetry.
By construction the rotational symmetry is preserved.
Therefore, the use of an approximate initial field profile must destroy the boost symmetries.
An example of exactly this type of approximate intitial condition is to use the thin-wall approximation rather than the exact instanton solution.
Since the thin-wall approximation does not actually satisfy the instanton equation, this means it cannot respect the SO(3,1) symmetry.
When we obtain two bubble initial conditions by superposing two of the approximate initial conditions, the evolved fields will preserve the rotation symmetry about the collision axis but break the two boost symmetries.

\begin{figure}
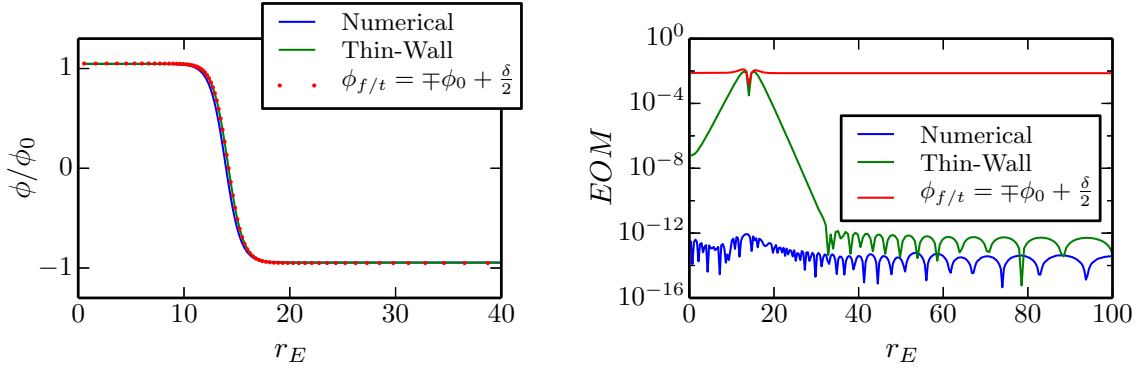

  \centering
  \includegraphics[width=0.48\linewidth]{{{instanton_linear_del0.1}}}
  \hfill
  \includegraphics[width=0.48\linewidth]{{{instanton_linear_del0.1_eom}}}
  \caption[Comparison of the exact instanton solution to various thin-wall approximations for $\delta=0.1$ in the linear potential~\eqref{eqn:potential_linear}]{The instanton profile for the linear symmetry breaking potential~\eqref{eqn:potential_linear} with $\delta=0.1$.  We compare the exact instanton solution with various thin-wall approximations.  On the left we show the numerically generated instanton profile used to initialize the bubbles.  For comparison, we also include the thin-wall approximation, both in the case when we determine the vacua exactly~\eqref{eqn:thinwall_exactvac}, and in the case when we approximate the vacua to $\mathcal{O}(\delta^3)$~\eqref{eqn:thinwall_approxvac}.  In the right panel we plot the violation of the instanton equation~\eqref{eqn:instanton} for both our numerical profile and the thin-wall approximations.  
When measured by the maximal violation of the equation of motion, our solution is ten orders of magnitude more accurate than the thin-wall result.  The mapping parameters were $d=0.35$ and $L=1.6R_{0}$.}
  \label{fig:compare_instantons}
\end{figure}

Here we consider the effect of using the thin-wall approximation combined with an approximate determination of $\phi_{\mathrm{false}}$ and $\phi_{\mathrm{true}}$.
This is a common approximation and is obtained if one writes the potential as $V_{sym}+\Delta V$ and solves for the profile of the wall by dropping the $\Delta V$ term (provided of course that the locations of the minima are perturbed by $\Delta V$).
Since we want the bubble profile to interpolate between the false and true vacua, we take the following initial condition for a bubble centered at ${\bf x_0}$
\begin{equation}
  \phi_{\mathrm{init}} = \frac{(\phi_f-\phi_t)}{2}\tanh\left(\frac{|{\bf x}-{\bf x_0}|-R_{init})}{w}\right) + \frac{\phi_f+\phi_t}{2}
  \label{eqn:thinwall_exactvac}
\end{equation}
with $mR_{\mathrm{init}} = \sqrt{2}\delta^{-1}$ and $mw=\sqrt{2}$.
The notation $\phi_{f/t}$ is meant to distinguish the approximate locations of the false and true vacua from the exact values $\phi_{false/true}$.
We further approximate the values of $\phi_f$ and $\phi_t$ to linear order in $\delta$.  We find ${\phi_{f} = -1+\delta/2+\mathcal{O}(\delta^3)}$ and ${\phi_{t}=1+\delta/2+\mathcal{O}(\delta^3)}$.
At this level of approximation, we have
\begin{equation}
  \phi_{\mathrm{init}} \approx \phi_0\tanh\left(\frac{(|{\bf x}-{\bf x_0}|-R_{init})}{w}\right) + \frac{\delta}{2} \, .
  \label{eqn:thinwall_approxvac}
\end{equation}
A comparison of approximation~\eqref{eqn:thinwall_approxvac}, the thin-wall approximation with an exact determination of $\phi_{false/true}$ and the numerical result are shown in~\figref{fig:compare_instantons}.
Clearly, the approximate solutions provide a very accurate description of the system initially,
although the perturbation to the initial bubble radius is visible in the figure.
A consequence of using approximate vacua is that the equation of motion is violated over the entire domain of $r_E$.

\begin{figure}
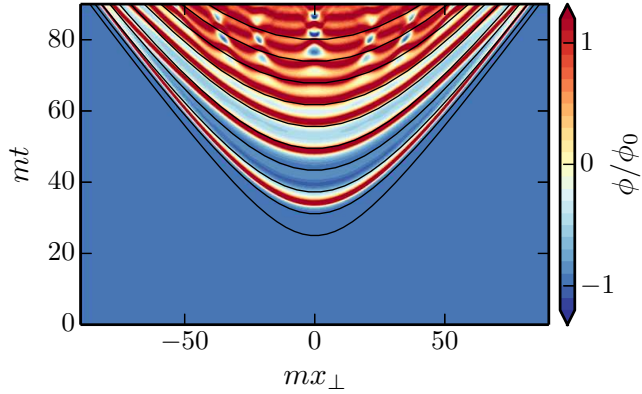

  \centering
  \includegraphics[width=0.6\linewidth]{{{bubble_badvac_transverse_del0.1}}}
  \caption[Field slice through the collision plane for a pair of colliding bubbles with initial conditions given by~\eqref{eqn:thinwall_approxvac}]{Field slice through the collision plane for a pair of colliding bubbles with initial conditions given by~\eqref{eqn:thinwall_approxvac} in the linear symmetry breaking potential with $\delta=0.1$.  The rotational symmetry about the collision axis is maintained to high-precision, but one can clearly see the loss of boost invariance as the system evolves.  The breaking of the boost symmetry develops on constant $s$ slices exactly as one expects for resonantly amplified fluctuations with $\chi$ dependence.}
  \label{fig:transverse_plane_badvac}
\end{figure}

\begin{figure}
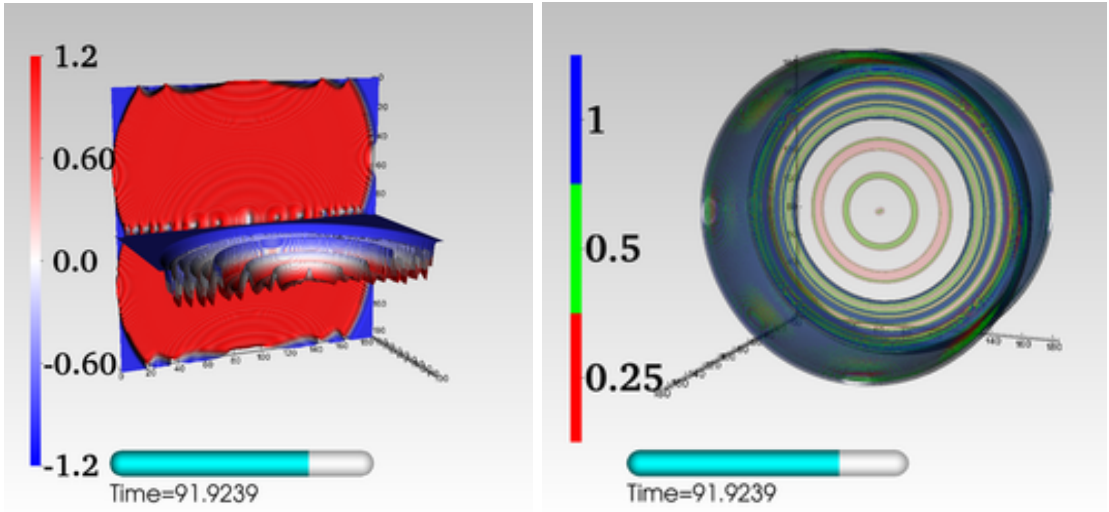

  \begin{center}
    \includegraphics[width=0.45\linewidth]{{{bubble_field_twowell_thinwall_0003}}} 
    \includegraphics[width=0.48\linewidth]{{{bubble_contour_twowell_thinwall_0003}}}
  \end{center}
  \caption[Same view of the field as~\figref{fig:transverse_slice_nofluc}, except for an initial bubble profile~\eqref{eqn:thinwall_approxvac} rather than the numerical result, and contours of energy density $T^{00}$]{\emph{Left}: The same view of the field as~\figref{fig:transverse_slice_nofluc}, except for an initial bubble profile~\eqref{eqn:thinwall_approxvac} rather than the numerical result.  \emph{Right}: Contours of energy density $-\frac{T^{0}_0}{\lambda\phi_0^4}$ for the same simulation.  A comparison with~\figref{fig:transverse_slice_nofluc} shows that the collision results in the complete destruction of the boost symmetry.  An animation of the field evolution is available at \thinwallapproxfield.}
  \label{fig:transverse_slice_taufluc}
\end{figure}

\begin{figure}
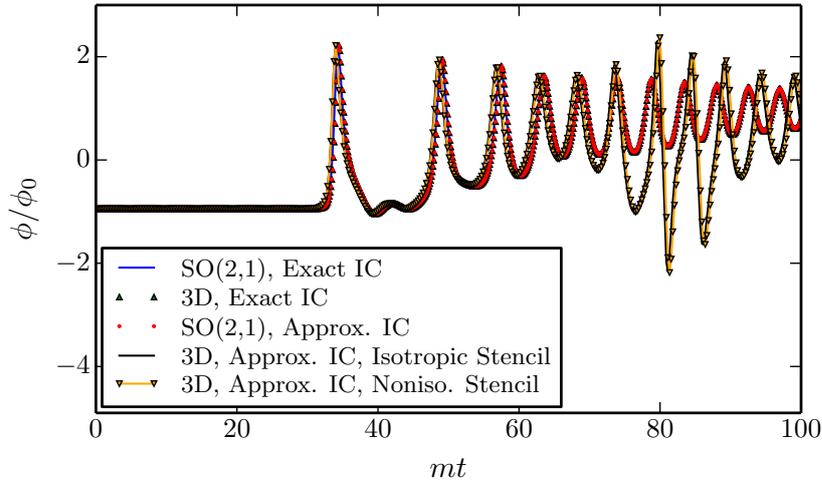

  \centering
  \includegraphics[width=0.75\linewidth]{{{field_origin_linear_del0.1_wbadvac}}}
  \caption[Direct comparison between 1D simulations using a 10th order Gauss-Legendre integrator and 3D simulation with Fourier pseudospectral approximation for spatial derivates]{Direct comparison between our one-dimensional simulations using a 10th order Gauss-Legendre integrator and Fourier pseudospectral approximation for spatial derivatives, and our full three-dimensional parallel lattice simulations.  
We plot the value of the field at the initial point of collision between the two bubbles as a function of time. 
If we use the correct initial instanton profile the two results agree extremely well, apart from a small shift in the time of the initial collision.
However, once we break the boost symmetry by using approximate choices for the locations of the true and false vacuum large deviations between the solutions appear.  
This effect is \emph{not} captured by a one-dimensional lattice that assumes SO(2,1) symmetry for the evolving field.}
  \label{fig:thin_wall_badvac_center}
\end{figure}

From~\figref{fig:transverse_plane_badvac} we see that the symmetry breaking fluctuations grow and become nonlinear on fixed $s$ slices,
 exactly as expected for perturbations associated with the $\chi$ direction.
To check this is not a numerical artifact, we ran the same simulation using two different choices for the finite-difference stencil.
In all cases the resulting evolution was very similar as seen in~\figref{fig:thin_wall_badvac_center}.
This, coupled with the preservation of the symmetry for the exact instanton initial condition and the explanation of the linear instability in terms of parametric resonance~\cite{ref:bbm1}, strongly indicates that the boost symmetry breaking is a real effect and not an artifact of the the discrete lattice spacing and time steps.
Note that the strong breaking of the boost symmetry is absent in the solution to the dimensionally reduced dynamics~\eqref{eqn:bubble_bg}. 
Thus, this is an effect which is only captured by intrinsically higher-dimensional simulations (in this case three-dimensional).

\section{Bubble Collisions with Fluctuations}
\label{sec:one_field}
In section~\ref{sec:symmetric_collision} we verified the accuracy of our lattice code and demonstrated the need for a full three-dimensional treatment of the two-bubble problem even in a simple case where we break the boost symmetries but not the rotational symmetry.
We now include a complete set of \emph{bulk} fluctuations that break both the rotational symmetries and the boost symmetries.
As we demonstrate shortly, the dynamical amplification of these fluctuations can lead to a complete breakdown of all of the boost and rotational symmetries in the system.
This section constitutes the core part of our analysis.
We consider several classes of single-field potentials, and
we find that the extreme breaking of the spacetime symmetries is restricted to double-well potentials with mildly broken $Z_2$ symmetry.
Section~\ref{sec:two_field} extends the analysis to a two-field case.

From our experience with the boost breaking fluctuations in section~\ref{sec:boost_breaking} and the linear analysis performed in~\cite{ref:bbm1}, it is clear that asymmetric fluctuations will undergo resonant amplification for certain types of bubble collisions.
Eventually the amplified fluctuations will begin to interact nonlinearly.
At this point the split between background and fluctuations becomes blurred and we must study the full \emph{three-dimensional} nonlinear field theory.
For a planar symmetric wall-antiwall pair, we demonstrated in~\cite{ref:bbm2} that the amplification of fluctuations transverse to the collision axis eventually lead to the dissolution of the walls and the creation of a population of oscillons distributed within a narrow transverse slab centered the site of the collision.
Since we are considering bubbles here, the ``transverse'' wavenumbers can be split into a wavenumber associated with the radial hyperbolic direction $\chi$ and one associated with rotation about the collision axis $\theta$.  For definitions of $\chi$ and $\theta$ see~\eqref{eqn:so2_coordinates}.
The fluctuations with $\chi$ dependence only were studied in section~\ref{sec:boost_breaking} where we looked at boost symmetry breaking fluctuations.
In the remainder of the paper we explore the full evolution of three-dimensional bubble collisions with small initial fluctuations around the instanton profiles for a variety of potentials.
In particular, we study the final outcome of the nonlinear interactions of the (linearly) amplified fluctuations.
For cases where the bubble walls bounce off of each other many times, we will see that the results match our intuition developed from the planar wall limit studied in~\cite{ref:bbm2}.

We take the initial bubble separation and the initial amplitude of the fluctuations to be independent free parameters.
However, if we wish to study collisions between typical bubbles this will not be true.
The action of the bounce, which determines the nucleation rate and therefore the typical bubble separation, scales as $\lambda^{-1}$.
Meanwhile, the RMS amplitude of the fluctuations scales as $\sqrt{\lambda}$.
Therefore, increasing the amplitude of the fluctuations has the effect of increasing the nucleation rate and thus decreasing the typical bubble separation.
Our primary motivation to treat these parameters independently was numerical.
We simulate a finite volume with periodic boundary conditions, and therefore we have a finite amount of time before the bubbles begin to interact with their images and we are no longer simulating a simple two bubble collision.
To study the effects of nonlinear interactions, the exponential growth of the fluctuations must push them into the nonlinear regime before the effects of periodicity reach the collision region.
We chose the initial amplitude (while maintaining $\frac{\delta\phi}{\phi_0} \ll 1$) to ensure this was the case.
Of course, since the nucleation events are stochastic, for a given model a range of separations between nucleated bubbles will occur.

\subsection{Thin-Wall Double Well Case}
\begin{figure}[h]
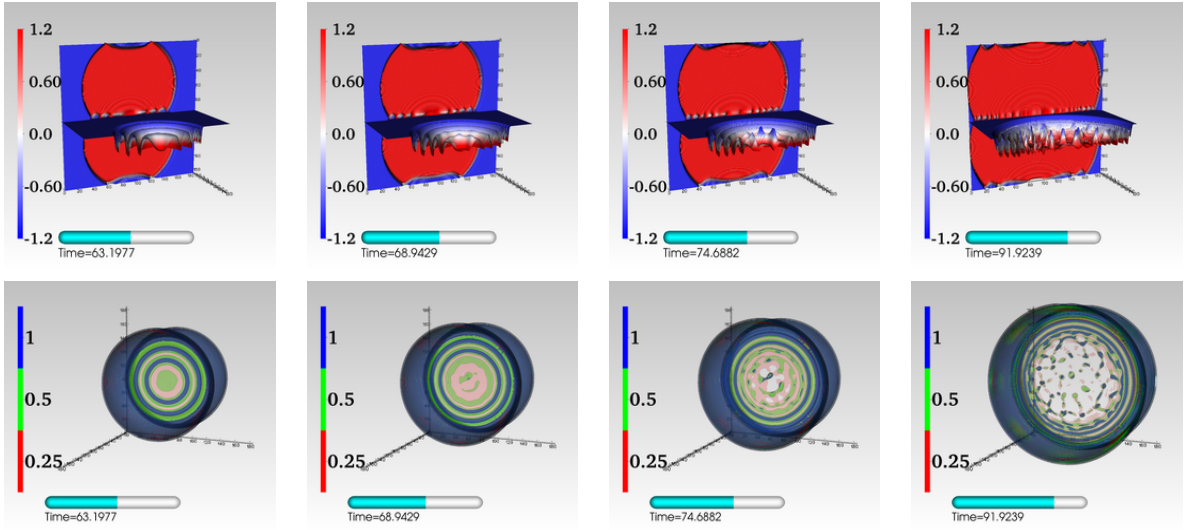

  \begin{center}
  \begin{tabular}{cccc}

  \includegraphics[width=0.23\linewidth]{{{bubble_field_twowell_wfluc_0001}}}&
  \includegraphics[width=0.23\linewidth]{{{bubble_field_twowell_wfluc_0002}}}&
  \includegraphics[width=0.23\linewidth]{{{bubble_field_twowell_wfluc_0003}}}&
  \includegraphics[width=0.23\linewidth]{{{bubble_field_twowell_wfluc_0006}}}\\

  \includegraphics[width=0.23\linewidth]{{{bubble_contour_twowell_wfluc_0001}}} &
  \includegraphics[width=0.23\linewidth]{{{bubble_contour_twowell_wfluc_0002}}} &
  \includegraphics[width=0.23\linewidth]{{{bubble_contour_twowell_wfluc_0003}}} &
  \includegraphics[width=0.23\linewidth]{{{bubble_contour_twowell_wfluc_0006}}} 

  \end{tabular}
  \caption[Development of the instability for two colliding bubbles in Minkowski space including bulk fluctuations corresponding to $\lambda = 10^{-4}$ around the thin-wall approximation]{Development of the instability for two colliding bubbles in Minkowski space including bulk fluctuations corresponding to $\lambda = 10^{-4}$ around the thin-wall approximation.  In the top row we plot the field distribution on a 2-D slice through the plane where the bubbles collide (\emph{horizontal projection}) and on a slice parallel to the collision axis through the centers of the bubbles (\emph{vertical projection}).  In the bottom row we show contours of the energy density.  Our lattice has $N=1024$ points per side with a box size of $mL = 13R_0 = 13\sqrt{2}\delta^{-1}$ and a time step $dt = dx / 12.5$.  The bubbles are nucleated with their centers separated by a distance $mR_{sep} = \frac{26}{5}R_{0}$.  Videos corresponding to both sets of figures can be found at \thinwallflucfield\ and \thinwallrho.}
  \label{fig:bubble_collision_thinwall_fluc}
  \end{center}
\end{figure}

Our first example is the collision of two thin-walled bubbles in the linear symmetry breaking potential with $\delta=0.1$.
Aside from the new scale associated with the radius of the bubbles, 
this case is qualitatively the same as the collision of a pair of planar walls in the same potential.
In section~\ref{sec:boost_breaking}, we showed that perturbing the bounce solution slightly can lead to a dramatic breaking of the two boost symmetries.
From our previous study of planar walls~\cite{ref:bbm2},
we expect that the evolution with a full complement of initial fluctuations will ultimately lead to the dissolution of the bubble walls and the creation of a collection of oscillons in the collision region.
\Figref{fig:bubble_collision_thinwall_fluc} demonstrates that this is indeed the case.
Near the centre of the collision, the bouncing of the walls amplifies the (initially small) fluctuations.
At $mt \sim 40$, these fluctuations become of similar amplitude to the oscillations of the SO(2,1) background.
Shortly after this, the distinction between background and fluctuations in the collision region breaks down and the field rapidly condenses into a population of oscillons.
In our choice of time coordinate, the condensation occurs first near the center of the collision.
Meanwhile, the outward propagating torii produced during the first few collisions of the walls develop ripples that eventually pinch off leading to the creation of ``rings'' of oscillons near the outer edges of the collision region.
The two mechanisms described above correspond to the two production mechanisms we anticipated in section~\ref{sec:bubbles_linear}:
parametric amplification of fluctuations by the oscillating background field near the collision center, and the growth of fluctuations on the outward propagating torii.

\subsection{Thin-Wall with Plateau}
\begin{figure}[!ht]
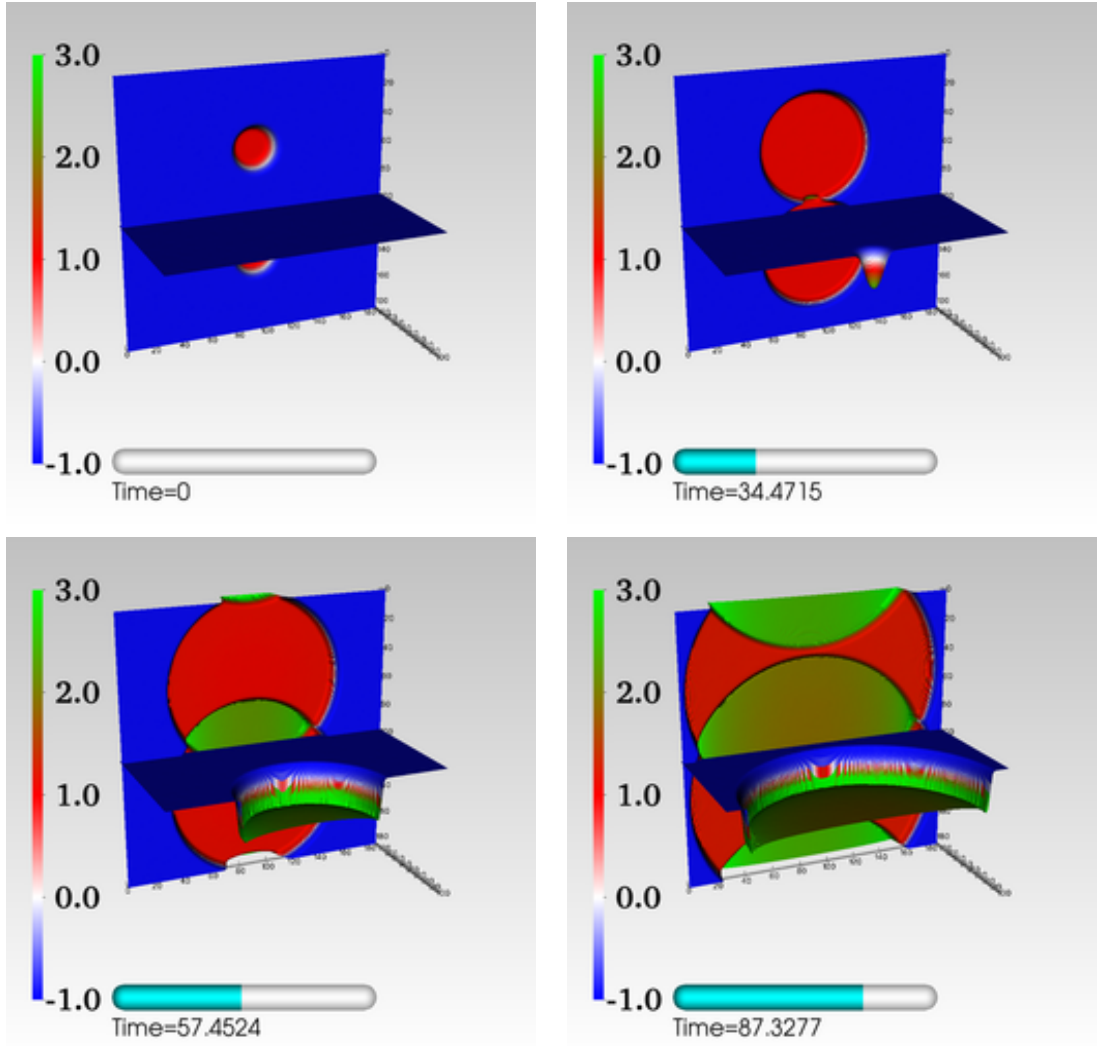

  \begin{tabular}{cc}
  \includegraphics[width=0.45\linewidth]{{{bubble_plateau_0000}}} &
  \includegraphics[width=0.45\linewidth]{{{bubble_plateau_0001}}} \\
  \includegraphics[width=0.45\linewidth]{{{bubble_plateau_0002}}} &
  \includegraphics[width=0.45\linewidth]{{{bubble_plateau_0003}}}
  \end{tabular}
  \caption[A collision of two single-field bubbles with a flat plateau onto which the field tunnels rather than a second well]{A collision of two single-field bubbles with a flat plateau onto which the field tunnels~\eqref{eqn:onefield_plateau}.  The horizontal field slices are orthogonal to the collision axis and through the center of the collision region. For orientation, the back panel is a slice parallel to the collision axis, illustrating the growth of the two bubbles.  Blue corresponds to regions where the field is near the false vacuum, red regions where it is near the tunnel out point, and green regions down the plateau away from the tunnel out point.  A illustrative video of the process can be found at~\plateaufield.}
  \label{fig:plateau_field}
\end{figure}

For our next example potential, we modify the region into which the field tunnels by inserting a long-flat plateau rather than a second well.
This new potential is given by
\begin{displaymath}
  \label{eqn:onefield_plateau}
  V(\phi) = \left\{
    \begin{array}{lr}
      \frac{1}{4}\left(\phi^2-\phi_0^2\right)^2 - \delta\phi_0^3\phi +\tilde{V}_0   & : \phi < \phi_{true} \\
      V_0 - \epsilon(\phi-\phi_{true})& : \phi > \phi_{true}
    \end{array}
  \right.
\end{displaymath}
where $V_0 = \frac{1}{4}(\phi_{true}^2-\phi_0^2)^2 - \delta\phi_0^3\phi_{true} + \tilde{V}_0$ and (assuming $\delta > 0$) $\phi_{true}$ is the largest solution to $\phi_{true}^3-\phi_{true}\phi_0^2-\delta\phi_0^3=0$.
This is meant as a toy example where the field tunnels out onto a flat inflationary plateau, with the slope determined by $\epsilon \geq 0$.
The portion of the potential traversed by the instanton as it tunnels is unchanged from the double well case~\eqref{eqn:potential_linear}, and therefore the initial bubble profile is the same.
However, we expect the collision dynamics between this case and the thin-walled double-well to be radically different.
For a sufficiently energetic collision, the free passage approximation~\cite{Easther:2009ft,Giblin:2010bd,Amin:2013dqa,Amin:2013eqa} will hold shortly after collision.
This tells us that the collision will displace the field a distance $\Delta\phi \sim 2\phi_0$ down the plateau.
Unlike the double-well, this does not result in a restoring force pulling the field back towards the false vacuum.
Therefore, there is no bouncing of the walls or other oscillations of the background field.
Thus, there is no mechanism to pump fluctuations. 
Rather, the effect of the collision is to create a field gradient interpolating between the field value at the tunnel out location
and its displaced value down the plateau.
This gradient then propagates away from the collision as seen in~\figref{fig:plateau_field}.
In front of the gradient, the field is at $\phi \approx \phi_{true} \approx \phi_0$ while behind it the field has been displaced down the plateau to $\phi \approx \phi_{true}+2\phi_0 \approx 3\phi_0$.
Although this behaviour is very similar to that of a propagating domain wall, the gradient does \emph{not} interpolate between local minima of the potential.
Thus, unlike a domain wall there are no bound state fluctuations associated with the interpolating field because $V''$ is never negative.
Somewhat suprisingly, the field at the collision site does not remain stationary on the plateau.
Rather, it begins to slowly retreat back to the tunnel out location, at least for the duration of our simulations.
From~\figref{fig:plateau_compare}, we see that the field profiles both with and without initial fluctuations are nearly the same,
indicating that the fluctuations play a subdominant role for this choice of potential.
As a result, this is a situation in which the SO(2,1) assumption provides an accurate description of the evolution.

Now consider the effect of adding a nonzero slope to the flat plateau.
Inside the bubble the field begins to roll down the plateau once it tunnels.
Again, the collision effectively displaces the field further down the plateau and produces a steep gradient which propagates into the bulk.
This gradient propagates on top of the previous rolling of the field down the potential.
As with the flat plateau, \figref{fig:plateau_compare} shows that the resulting evolution is insensitive to the choice of whether or not we include fluctuations.
Therefore, we expect this situation to also be well described by 1+1-dimensional simulations.
\begin{figure}[!ht]
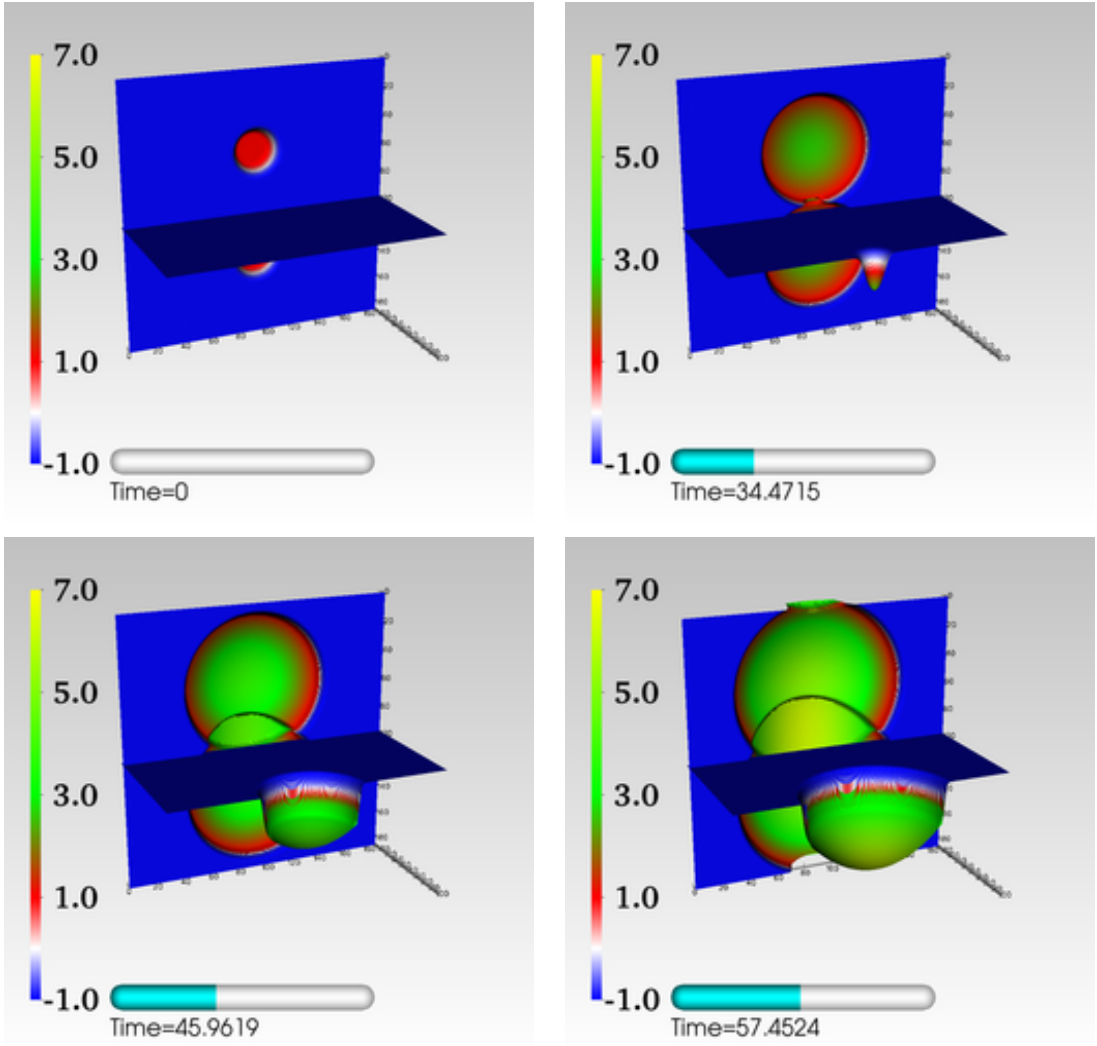

  \begin{center}
    \begin{tabular}{cc}
      \includegraphics[width=0.45\linewidth]{{{bubble_plateau_tilt_0000}}} &
      \includegraphics[width=0.45\linewidth]{{{bubble_plateau_tilt_0001}}} \\
      \includegraphics[width=0.45\linewidth]{{{bubble_plateau_tilt_0002}}} &
      \includegraphics[width=0.45\linewidth]{{{bubble_plateau_tilt_0003}}}
    \end{tabular}
  \end{center}
  \caption[A collision of two single-field bubbles with a tilted plateau $\epsilon=0.01$]{A collision of two single-field bubbles with a tilted plateau $\epsilon=0.01$.  The field slices are the same as in~\figref{fig:plateau_field}.
Inside the bubble, the field now rolls down the potential.
However, the result of the collision is essentially the same as in the case $\epsilon=0$.
The field is displaced down the potential at the collision, and a field gradient then propagates into the interior of the bubble.
Aside from this, the field rolls down the potential due to the constant force resulting from the slope.}
  \label{fig:plateau_field_tilt}
\end{figure}

\begin{figure}[h]
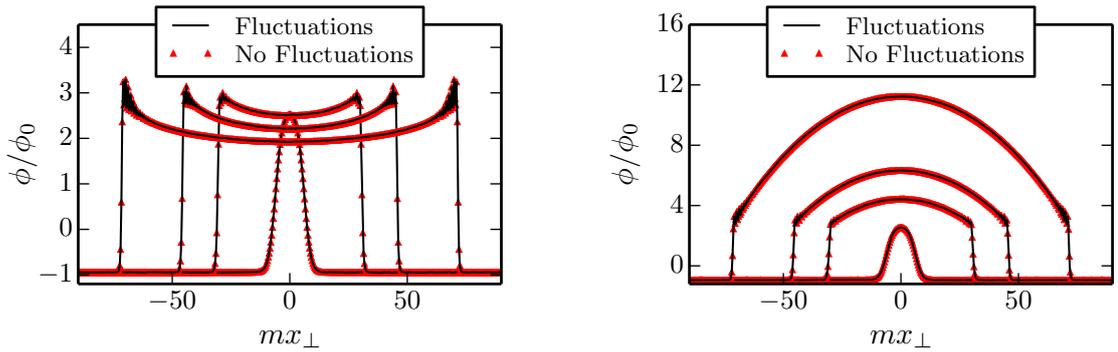

  \centering
  \includegraphics[width=0.48\linewidth]{{{plateau_compare_transverse}}}
  \hfill
  \includegraphics[width=0.48\linewidth]{{{plateau_tilt_compare_transverse}}}
  \caption[Comparison of bubble profiles orthogonal to the collision axis for a simulation with initial fluctuations and without initial fluctuations $\delta\phi = 0$]{Comparison of bubble profiles orthogonal to the collision axis for a simulation with initial fluctuations $\delta\phi \neq 0$ (\emph{black line}) and without initial fluctuations $\delta\phi = 0$ (\emph{red triangles}).  The choice of potential and initial conditions are the same as \figref{fig:plateau_field} (\emph{left}) and \figref{fig:plateau_field_tilt} (\emph{right}).  We plot four times $mt=34.47$, $mt=45.96$, $mt=57.45$ and $mt=80.43$.  The width of the region displace from the false vacuum ($\phi=-\phi_0$) increases with time.  The two profiles match extremely well at all times, indicating that the fluctuations are stable for this class of bubble collisions.  The oscillations near the edge of collision region are numerical artifacts.  In the left panel we show the result for a flat plateau $\epsilon = 0$, while on the right we have $\epsilon = 0.01$.  For the purposes of plotting, we have downsampled our output grid by a factor of 2.}
  \label{fig:plateau_compare}
\end{figure}

\subsection{Thick-Wall Double Well Case}
\begin{figure}[h]
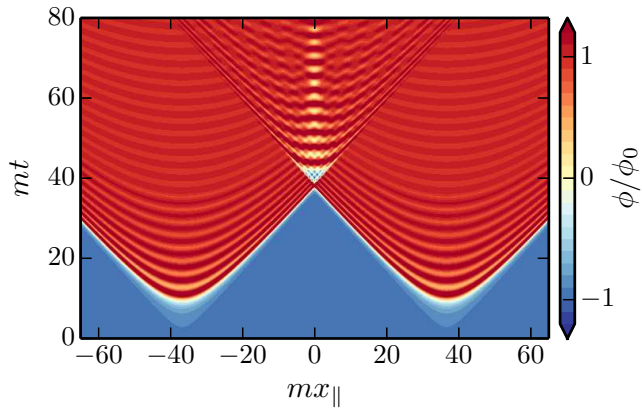

  \begin{center}
    \includegraphics[width=0.6\linewidth]{{{two_bubble_1d_del0.99_collision}}}
  \end{center}
  \caption[Evolution of a pair of colliding bubbles in the cubic potential with $\delta=0.99$ under the assumption of exact SO(2,1) symmetry]{The evolution of a pair of colliding bubbles in the cubic potential with $\delta=0.99$ under the assumption of exact SO(2,1) symmetry.  We plot the field as a function of time and position along the collision axis.}
  \label{fig:bg_collision_thickwall}
\end{figure}

\begin{figure}[h]
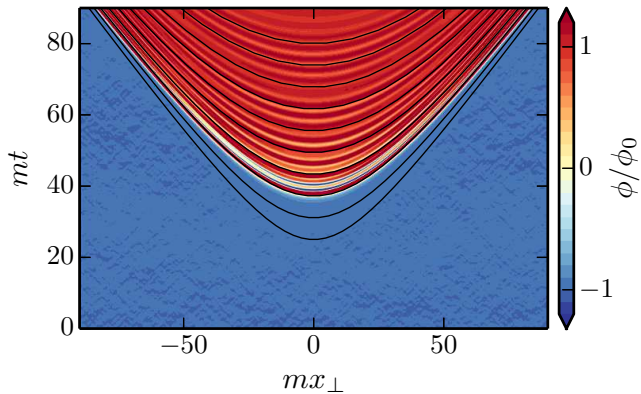

  \centering
  \includegraphics[width=0.6\linewidth]{{{twobubble_thickwall_transverse}}}
  \caption[Evolution on a line orthogonal to collision axis showing the preservation of the boost symmetry]{Evolution on a line orthogonal to collision axis through the initial point of collision for a collision between thick-walled bubbles in the cubic symmetry breaking potential with $\delta=0.99$.  We have taken a fairly large initial fluctuation amplitude to demonstrate the preservation of the boost symmetry.  The fluctuations were those in the Minkowski false vacuum with $\lambda = 10^{-2}$.}
  \label{fig:thick_wall_collision_compare}
\end{figure}

As our final potential we consider a collision between two thick-walled bubbles in the cubic symmetry breaking potential with $\delta=0.99$.
In~\figref{fig:bg_collision_thickwall} we show the result of one such collision, where we have made the assumption of SO(2,1) symmetry in order to run a very high resolution one-dimensional simulation.
The collision still leads to large oscillations of the field around the minimum, but the the field no longer becomes temporarily trapped in the false vacuum minimum between collisions.
Instead of having repeated collisions between a pair of walls, we instead have oscillations of the field around the true vacuum.
Within a few oscillations, the amplitude damps to $\lesssim 0.5\phi_0$.
We already know from our study of single field bubbles that spatially homogeneous oscillations of the field with this amplitude do not lead to a strong instability.
Since the oscillations from the collision have additional spatial localization, it is clear that the fluctuations will not experience significant amplification in this case either.
We see this directly in~\figref{fig:thick_wall_collision}, where the rotational symmetry about the collision axis is preserved to a high degree during the collision.
As well,~\figref{fig:thick_wall_collision_compare} further shows that the boost symmetry is extremely well preserved.
\begin{figure}[!ht]
  \begin{center}
  \includegraphics[width=0.45\linewidth]{{{bubble_field_twowell_thickwall_0000}}} 
  \includegraphics[width=0.45\linewidth]{{{bubble_field_twowell_thickwall_0001}}} \\
  \includegraphics[width=0.45\linewidth]{{{bubble_field_twowell_thickwall_0002}}}
  \includegraphics[width=0.45\linewidth]{{{bubble_field_twowell_thickwall_0003}}}
  \end{center}
  \caption[A collision of two-thick wall bubbles in the potential~\eqref{eqn:potential_cubic} with $\delta = 0.99$]{A collision of two-thick wall bubbles in the potential~\eqref{eqn:potential_cubic} with $\delta = 0.99$.  The field slices and coloring are the same as in~\figref{fig:bubble_collision_thinwall_fluc}.  A video of the field evolution is available at \thickwallfield.  } 
  \label{fig:thick_wall_collision}
\end{figure}

\subsection{Bubble-Domain Wall Collision}
Thus far we have only considered the collision of two bubbles in a reference frame in which they nucleate at the same time.
We will now address the opposite regime where one of the bubbles is much larger than the other.
To model this situation we consider the collision of a bubble with a planar domain wall, with the planar wall meant to be a substitute for the large bubble.
In an actual collision between false vacuum bubbles, the wall has a very large Lorentz contraction factor $\gamma \gg 1$.
This creates a very large hierarchy between the size of the smaller bubble and the thickness of the planar wall which we are unable to resolve with a fixed lattice.
To avoid this problem we restrict ourselves to mildly relativistic walls.
A physical situation where this occurs is finite-temperature phase transitions,
where interactions between the bubble wall and the surrounding thermal medium prevent the speed of the bubble walls from nearing the speed of light.
Of course the initial spectrum of fluctuations is different in the thermal case and interactions with the surrounding thermal bath must also be accounted for.
Therefore a direct comparison with our results cannot be made, although the same qualitative behaviour will persist.
As well, we consider only collisions in the linear symmetry breaking potential~\eqref{eqn:potential_linear}, again with $\delta=0.1$.
\begin{figure}[h]
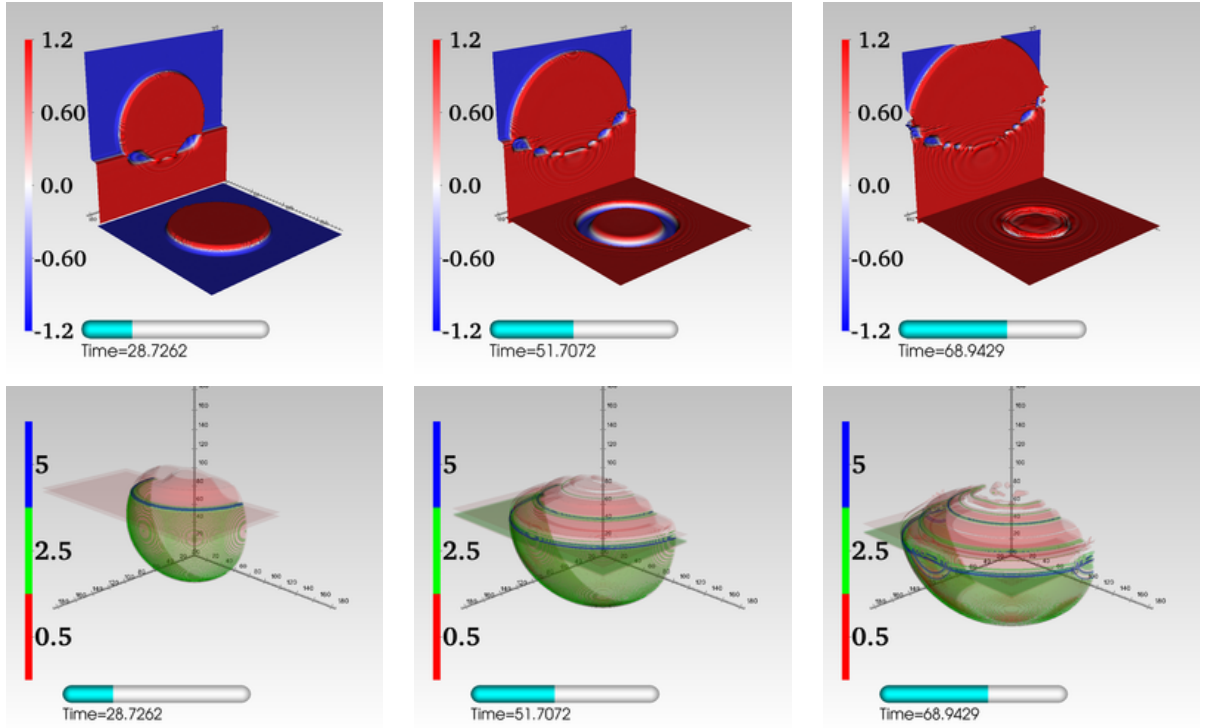

  \centering
  \begin{tabular}{ccc}
    \includegraphics[width=0.32\linewidth]{{{field_slices_bubble_wall_0001}}} &
    \includegraphics[width=0.32\linewidth]{{{field_slices_bubble_wall_0002}}} &
    \includegraphics[width=0.32\linewidth]{{{field_slices_bubble_wall_0003}}} \\

    \includegraphics[width=0.32\linewidth]{{{density_contours_bubble_wall_0000}}} &
    \includegraphics[width=0.32\linewidth]{{{density_contours_bubble_wall_0001}}} &
    \includegraphics[width=0.32\linewidth]{{{density_contours_bubble_wall_0002}}}
  \end{tabular}
  \caption[Evolution for a collision between a planar domain wall and a single vacuum bubble.]{Field and energy density evolution for a collision between a planar domain wall and a single nucleated bubble.  In the top row we show two slices of the field configuration.  The first slice (projected vertically) is perpendicular to the planar wall and cuts through the center of the bubble.  The second slice (projected horizontally) is parallel to the domain wall and displaced slightly from the center of the bubble towards the initial location of the domain wall.
    As the wall sweeps past the bubble, the instantaneous collision location moves along the collision axis, as do the tubes of false vacuum and oscillons produced by the collision.  In the bottom row, we instead show contours of energy density for the same three time slices.}
  \label{fig:bubble_and_wall}
\end{figure}
From our results for the collision of two bubbles nucleated at rest at the same time, it is clear that these types of collisions will again result in strong amplification of the fluctuations and the eventual production of a population of oscillons.
The only new ingredient here is that the planar wall and bubble wall no longer carry an equal and opposite amount of field momentum density along the collision axis $\mathcal{P}^{x_\parallel} = \dot{\phi}\partial^{x_\parallel}\phi$.
Therefore, if we consider a small box around any region where the planar wall and bubble are undergoing a collision, the energy within that box carries a nonzero momentum along the collision axis.
As well, the collision no longer occurs in a single spatial plane, but rather on a curved hypersurface defined by all of the instantaneous intersections between the bubble and the planar wall.
As a result, the collision products are produced along a nonplanar surface with a nonzero velocity (both relative to our simulation coordinates).
Since we are not properly boosting an initial two-bubble field configuration to set the initial conditions, the collision hypersurface will be nonplanar in any reference frame.

Figure~\ref{fig:bubble_and_wall} shows that these expectations are true.  As the wall accelerates into the false vacuum it partially engulfs the expanding bubble.
As with the two bubble case, the field experiences a series of excursions back to the false vacuum.
These drive an instability in linear (non rotationally invariant) fluctuations analogous to the instability in the pair of bubbles.
The torii resulting from these excursions no longer distribute themselves in a single plane, but rather they move in the same direction as the wall.
As a result, the oscillons that are ultimately produced from these growing fluctuations are distributed in some thin curved hypersurface rather than a plane.

\section{Extension to Two Fields}
\label{sec:two_field}
By considering the simple case of a single-field potential, 
we discovered that the dynamics of bubble collisions can be considerably more intricate than expectations based on $SO(2,1)$ or $SO(2)$ symmetry for the field profile.
In particular, for single-field double-well potentials with moderately broken $Z_2$ symmetry we found that the collision of a pair of bubbles ultimately leads to the production of a population of oscillons.
These types of symmetry breaking potentials are often used to model first-order phase transitions, and thus our results may have some relevance in that context.
However, the bubbles must coalesce in order for the phase transition to complete.
When the bubbles begin to coalesce we must consider collisions between many bubbles rather than just two, leading to an explicit and large breaking of the SO(2,1) symmetry.

A natural scenario to consider collisions between isolated pairs of bubbles is in false vacuum eternal inflation.
Unfortunately, it seems exceedingly difficult to both realize inflation and produce oscillons in bubble collisions using a single-field potential with only two minima.
The simplest way to have a viable single field open inflationary model produced by the nucleation of a bubble is to have the field tunnel out onto a plateau that can support $\Delta\ln a \sim 50$ efolds of inflation.\footnote{Another possibility is that collision displaces the field to a location where it begins to slow-roll, thus starting inflation.}
Our mechanism for oscillon production relies on oscillations of the field around the true vacuum within the collision region.
However, if the plateau is long enough to support a sufficient period of inflation, 
any rolling motion of the field along the plateau must experience many efolds worth of Hubble overdamping.
Therefore, the requirement of an inflationary plateau makes it difficult for the field to rebound off of a 
``wall'' in the potential and return to the false vacuum side of the barrier shortly after collision.
Thus the basic mechanism by which initial fluctuations are amplified disappears.

However, embedding inflationary models based on bubble nucleation into a realistic high-energy theory will likely involve considering models with many scalar fields.\footnote{Even this statement is probably too simplistic, as the correct high-energy theory may not even be describable in terms of a scalar field theory.}
With this in mind, consider a simple two-field potential
\begin{equation}
  V(\sigma,\phi) = \frac{\lambda_{\sigma}}{4}\left(\sigma^2 - \sigma_0^2\right)^2 + \lambda_\sigma\delta\left(\frac{\sigma^3\sigma_0}{3} -\sigma\sigma_0^3 + \frac{2\sigma_0^4}{3}\right) + \lambda_\sigma\frac{g^2}{2}\left(\sigma-\sigma_0\right)^2\phi^2 + \lambda_\sigma\sigma_0^3\epsilon\phi + V_0 \, .
  \label{eqn:potential_2field}
\end{equation}
Schematically, the potential has the form $V_{tunnel}(\sigma) + V_{coupling}(\sigma,\phi) + V_{inflation}(\phi)$, where we have chosen the particular form $V_{inflation}(\phi) = \lambda_\sigma\sigma_0^3\epsilon\phi + V_0$.
Our choice $V_{inflation} = \lambda_\sigma\sigma_0^3\epsilon\phi + V_0$ can be viewed as a linearization of the potential around the tunnel out region.
There is a local minimum at $\sigma\approx -\sigma_0$ and $\phi\approx 0$, and a long trough at $\sigma \approx \sigma_0$ along which $\sigma$ is heavy and $\phi$ is light.
Since we now have two mass scales at our disposal, we can accomodate the production of oscillons by exciting the $\sigma$ field while simultaneously permitting slow-roll inflation along the $\phi$ direction.
By adjusting $V_{inflation}$ we can effectively reproduce any model of single-field inflation we wish.
\begin{figure}
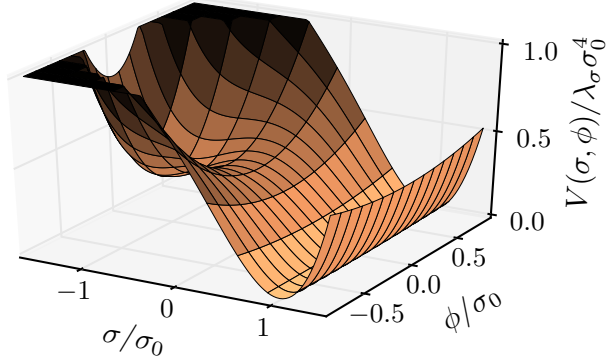

  \centering
  \includegraphics[width=0.6\linewidth]{{{potential_2field}}}
  \caption[Two-field potential for $\epsilon=0.01$, $\delta=0.2$, $g^2=1$ and $V_0=0$]{The two-field potential~\eqref{eqn:potential_2field} for $\epsilon=0.01$, $\delta=0.2$, $g^2=1$ and $V_0=0$.  
We have in mind a scenario where the field is initially trapped near the local minimum at $\sigma\approx -\sigma_0, \phi\approx 0$ and subsequently tunnels into the nearly flat trough at $\sigma=\sigma_0$.
For clarity we have clipped the potential for $V > \lambda\sigma_0^4$.}
\end{figure}

\begin{figure}
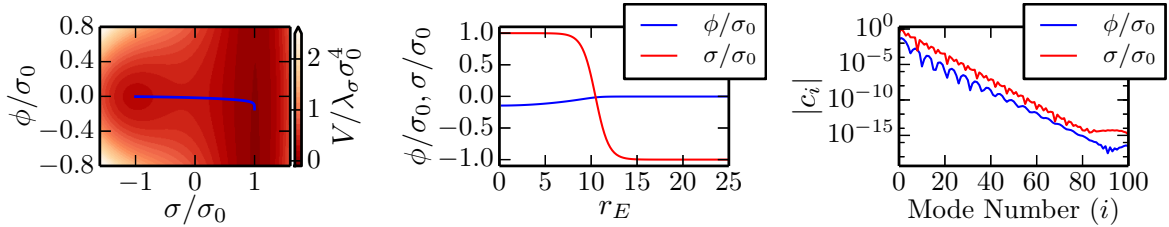

  \centering
  \includegraphics[width=0.32\linewidth]{{{2field_path_cubic_del0.2_eps-0.01}}}
  \hfill
  \includegraphics[width=0.32\linewidth]{{{2field_fields_cubic_del0.2_eps-0.01}}}
  \hfill
  \includegraphics[width=0.32\linewidth]{{{2field_spec_cubic_del0.2_eps-0.01}}}
  \caption[Instanton solution used to set initial conditions for our numerical simulation]{The instanton solution used to set initial conditions for our numerical simulation.  In the left panel we show the path of the instanton in field space superimposed on isocontours of the potential.  In the middle panel we show the profile of the tunnelling field $\sigma$ and the ``inflaton'' field $\phi$ as a function of the Euclidean radius $r_E$.  In the right panel we plot the spectral coefficients for each of the fields, from which we can see the exponential convergence of the series and the roundoff plateau arising around $i\sim 80$.  The model parameters were $\delta=1/5$, $\epsilon = 0.01$ and $g^2=1$.  Our mapping parameters were $L=1.6R_0^{cubic}$ and $d=0.5$, and we used 100 mode functions.}
  \label{fig:2field_instanton}
\end{figure}

\begin{figure}
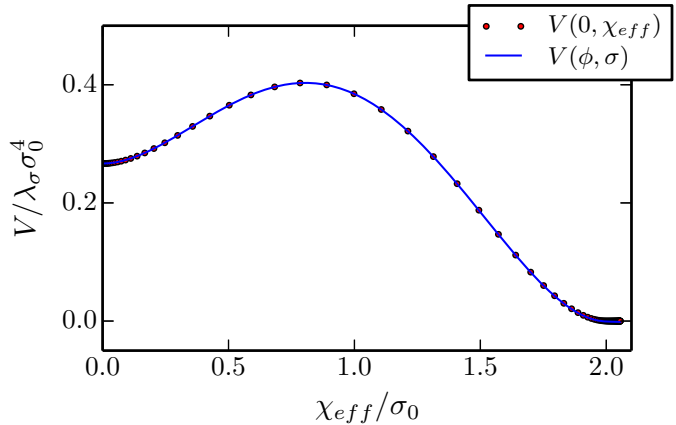

  \centering
  \includegraphics[width=0.6\linewidth]{{{sigma_eff_potential}}}
  \caption[Potential seen by the field as it tunnels in the Euclidean radial direction]{The potential seen by the field as it tunnels in the Euclidean radial direction (\emph{blue line}).  Here we have defined $d\chi_{eff}^2 = d\sigma^2 + d\phi^2$ as the path length in field space.  For comparison, we also include the potential $V(0,\chi_{eff})$ that would be seen by the field if we had instead tunnelled a distance $\chi_{eff}$ with $\phi=0$ (\emph{red dots}).  In this latter case, the potential is then the same as our single-field cubic symmetry breaking potential.}
  \label{fig:tunnelling_potential}
\end{figure}

We only consider parameter choices such that the tunnelling dynamics is dominated by $\sigma$, while the subsequent post-tunnelling evolution is dominated by $\phi$.
Although we have not performed an exhaustive study for different choices of $g^2$, when $g^2=1$ this behaviour is generic for the thin-walled case $\delta \ll 1$.
This is illustrated in~\figref{fig:2field_instanton}, where we explicitly see that during tunnelling the field first moves almost exclusively in the $\sigma$ direction
before making a sharp turn at the end so that it is moving along the slow-roll plateau. 
Effectively, the behaviour of the tunnelling field $\sigma$ is fixed by the double well structure, and the inflaton field $\phi$ simply reacts to the presence of the domain wall in $\sigma$.
In~\figref{fig:tunnelling_potential} we show the potential seen by the effective field $d\chi_{eff}^2 = d\sigma^2 + d\phi^2$ as it tunnels through the barrier,
as well as a comparison to an analogous single-field potential.
The structure of the potential as seen by the field while it tunnels and subsequently slow-rolls is very similar to the double-well with an appended slow-roll plateau~\eqref{eqn:onefield_plateau}.
As well, notice that there is no steepening of the potential near the beginning of inflation as is often assumed in single-field models of open inflation from bubble nucleation.

\begin{figure}
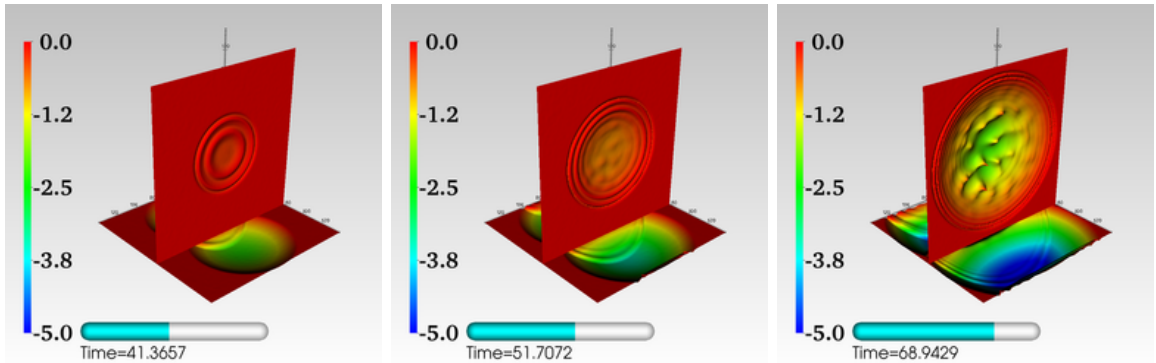

  \begin{center}
  \includegraphics[width=0.32\linewidth]{{{bubble_2field_inflaton_0001}}}
  \includegraphics[width=0.32\linewidth]{{{bubble_2field_inflaton_0002}}}
  \includegraphics[width=0.32\linewidth]{{{bubble_2field_inflaton_0003}}}
  \end{center}
  \caption[Evolution of the ``inflaton'' field $\phi$ in our two-field model]{Evolution of the ``inflaton'' field $\phi$ in our two-field model with $\epsilon=0.01$ and $\delta=0.2$.  The horizontal projection is a slice along the collision axis through the center of the bubbles and the vertical slice is orthogonal to the collision and centered on the collision site.  Red corresponds to $\phi$ values near the origin (\ie\ in the false vacuum or at the tunnel out location).  The pips where the field is pulled back up the potential are locations where the $\sigma$ field is fracturing into oscillons.  An animation corresponding to this evolution is available at \twofieldinflaton.}
  \label{fig:2field_inflaton}
\end{figure}

\begin{figure}
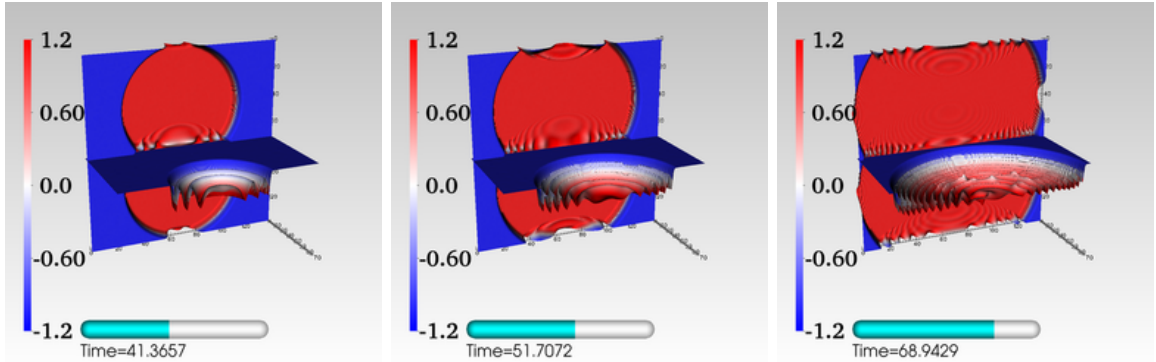

  \begin{center}
  \includegraphics[width=0.32\linewidth]{{{bubble_2field_tunnel_0001}}}
  \includegraphics[width=0.32\linewidth]{{{bubble_2field_tunnel_0002}}}
  \includegraphics[width=0.32\linewidth]{{{bubble_2field_tunnel_0003}}}
  \end{center}
  \caption[Evolution of the ``tunnelling'' field in our two-field model.]{Evolution of the ``tunnelling'' field $\sigma$ in our two-field model with $\epsilon = 0.01$ and $\delta = 0.2$.  As with previous figures, blue corresponds to $\sigma \sim -\sigma_0$ (\ie\ near the false vacuum) and red to $\sigma \sim \sigma_0$ (\ie\ in the ``inflationary'' trough).  The vertical slice is parallel to the collision axis and the horizontal slice is orthogonal to the collision axis.  An animation corresponding to this evolution is available at \twofieldtunnel.}
  \label{fig:2field_tunnel}
\end{figure}

\Figref{fig:2field_inflaton} and~\figref{fig:2field_tunnel} illustrate the collision dynamics between two of these multifield bubbles.
We do not perform an exhaustive study of possible behaviours as a function of initial bubble separation or model parameters.
Instead, we simply choose a set of parameters and initial conditions to demonstrate that oscillons can be produced in bubble collisions in this model.
For our choice of parameters, the collision causes a large excitation in $\sigma$ and its subsequent evolution is very similar to the single-field double well case.
However, for a fixed $\sigma_{cur} \not\approx \sigma_0$ the potential $V(\sigma_{cur},\phi)$ has a minimum at $\phi = -\epsilon\sigma_0^3 / (\sigma - \sigma_0)^2 \ll \sigma_0$.
As a result, the oscillations of $\sigma$ pull the field $\phi$ back towards the origin, thus undoing the previous rolling down the plateau.
The evolution of the $\sigma$ field amplifies fluctuations and leads to the creation of oscillons in the $\sigma$ direction.
This is analogous to the single-field evolution.
However, in the cores of the oscillons, the $\sigma$ field makes large excursions away from $\sigma_0$.
As a result, within these cores $\phi$ remains trapped at the origin while the field outside of these cores once continues to roll down the plateau.
In the inflationary setting, there will still be a vacuum energy $V_0$ sufficient to drive a period of slow-roll inflation,
so as the oscillons dilute we expect inflation to eventually restart.
A novel feature of this setup is that $\phi$ will be quite inhomogeneous (within the collision region) at the start of this new inflationary phase,
and we expect this inhomogeneity to persist until the oscillons decay.

\subsection{Inflationary Model Building and Consistency of the Minkowski Approximation}
Let's now consider the consistency of our approach and application of this type of model to inflation.
In our numerics, we have assumed that the instanton profile is well approximated by taking the background to be Minkowki.
As well, for our lattice simulations we have further assumed that we can account for expansion of the universe by taking a uniform fixed Hubble constant.\footnote{The figures presented in this paper were taken from simulations using a Minkowski background.  However, we also did several runs with a fixed Hubble constant $H$ over the entire simulation volume.  The main effect of this expansion was to delay the time that the bubbles first collide, while the collision dynamics itself was unaffected.}
Relaxing this latter assumption is a significant numerical challenge and we therefore leave it to future work.
This is required to obtain accurate predictions for the observational signatures that remain after inflation has occurred within the bubble.
While this is certainly an interesting question, in this paper we are primarily concerned with the dynamics of the collision itself.
The relevant time scale is then $\left(\lambda_\sigma\sigma_0^2\right)^{-1/2}$, and we assume this is much shorter than a Hubble time in the ambient spacetime in order for our approximations to be valid.

First consider the restrictions imposed by our computation of the instanton profiles.
The least stringent requirement is that the CdL instanton exists.
Roughly, this requires that the bubble wall fit within a Hubble radius determined at the local maximum of the potential $H_{max}^{-1}$.
For our setup, the thickness of the wall is determined by $V_{\sigma\sigma}(\sigma=0,\phi=0) \sim \lambda_\sigma \sigma_0^2$,
and therefore we require $ \lambda_\sigma \sigma_0^2 \gg H^2_{max} \sim V_0/3M_P^2 + \lambda_\sigma\sigma_0^4/12M_P^2 \implies 12\frac{M_P^2}{\sigma_0^2} \gg 1 + \frac{4V_0}{\lambda_\sigma\sigma_0^4}$.
A more stringent constraint comes from requiring that the initial radius of the bubble is much less than the Hubble scale in the ambient spacetime $H_{fv}^{-1}$.
We have $m_{eff}R_{init} \sim \delta^{-1}$ and $H_{fv}^2M_P^2 \sim \frac{4\delta\lambda\sigma_0^4}{9}+\frac{V_0}{3}$, so this gives $\frac{4\sigma_0^2}{3M_P^2}\left(1+\frac{3V_0}{4\delta\lambda_\sigma \sigma_0^4} \right) \ll \delta$.
In the limit that $V_0 \gg \delta\lambda_\sigma\sigma_0^4$, this gives $\frac{V_0}{\lambda_\sigma\sigma_0^2M_P^2} \ll \delta^2$.
For the opposite limit we instead get $\frac{\sigma_0^2}{M_P^2} \ll \delta$.

Finally, consider the restriction imposed by assuming the approximation of a fixed \emph{homogeneous} background Hubble.
This constraint only needs to be fulfilled if we want an approximate description of the dynamics prior to the collision,
or if we want to track the long time evolution after the collision.
The average expansion rate inside and outside the bubble must be much greater than the difference in these two expansion rates.
As well, for the case when the field slow-rolls inside of the bubble, our approximation will only be valid when the difference between the Hubble parameter in the center of the bubble (where the field has rolled furthest) and the edge is again much less than the average.
This then requires $V_0 \gg \delta\lambda_\sigma\sigma_0^4, |\lambda_\sigma\sigma_0^3\epsilon\phi_{{\bf x}=0}(t)|$ where $\phi_{{\bf x}=0}$ is the value of $\phi$ at the center of the bubble at time $t$.

\section{Observational Prospects}
\label{sec:observations}
In the previous sections we demonstrated the presence of a previously ignored instability in the fluctuations around colliding vacuum bubbles in double-well potentials.
As a result of these instabilities, the near SO(2,1) symmetry of the initial configuration becomes badly broken during the course of the dynamical evolution. 
This eventually results in the production of a population of oscillons in the collision region.
In this section we briefly comment on some possible implications of these results, restricting ourselves to potential signals that rely of the breakdown of the SO(2,1) symmetry.
This means we will not discuss possibilities such as a large scale modulation of coupling constants.
Modulations could result from single-field collisions in potentials with a plateau if the vev of the field forming the bubbles sets an effective coupling constant in the full theory.
There are many scenarios in which one can imagine embedding our mechanism, 
and we will distinguish between two cases that we refer to as superhorizon and subhorizon collisions.
In the first, we assume that our entire observable universe fits within one of the nucleated bubbles, with the nucleated bubble itself embedded in some parent false vacuum.  
The collision is then with a neighbouring bubble which has also nucleated within the parent vacuum.
Of course, this is the standard scenario for testing open inflation following bubble nucleation in an ambient false vacuum.
Past studies of signatures from such collisions are based on the assumption of SO(2,1) symmetry, so in this case in particular we would like to address any novel implications of the breakdown of the symmetry.
One could also imagine scenarios where our observable universe forms in the future light cone of the collision, although we will not discuss this possibility here.
In the second, we instead assume that the bubble nucleations are occurring on subhorizon scales within our Hubble volume.
These nucleations could either have occurred in the past (such as an early first order phase transition within our Hubble volume) or in the present. 
The reader should keep in mind that we have not performed a detailed analysis for any of the speculated possibilities in this section, so the magnitude of many of these effects may prove to be undetectably small.
We plan to provide a more detailed study in a future publication, including the effects of gravitation as required to obtain the present day signal.

\subsection{Superhorizon Bubble Collisions}
\label{sec:superhorizon_signatures}
First consider the case where our entire observable universe is contained within a nucleated bubble.
Our results demonstrate rather clearly that observational signatures of collisions do not necessarily posess an azimuthal symmetry about the collision axis as is assumed in the existing literature.
A remnant of the azimuthal symmetry persists in the sense that the effects of the collision are confined to a circular disk of the sky, but the interior of this disk need not posess any additional symmetries.

Perhaps the most interesting possibility is the production of gravitational waves by the fracturing of the bubble walls at the onset of nonlinearity amongst the fluctuations.
This effect is absent in an exact SO(2,1) collision, since the symmetries forbid the presence of tensor modes.
Initially the gravitational waves are produced with subhorizon sized wavelengths, but as the oscillons dilute and inflation restarts they are stretched outside the horizon.
The resulting signal will have a characteristic wavelength as well as a directional dependence on the sky that may be detectable in polarization data.
As well, the directional dependence of this polarization signal will be highly correlated with other possible signatures, for example a hot or cold spot produced by the collision.
Since the amplitude of the waves will damp until they are stretched outside of the horizon, the magnitude of the effect will depend on the typical scale of the nonlinear fracturing of the wall relative to the Hubble scale inside the bubble.

Another effect is related to sign of the temperature perturbation induced by the collision.
In the single-field case with the field tunnelling onto a plateau, the collision simply displaces the field down the plateau.
Therefore there are fewer efolds of inflation in the collision region than the rest of the bubble, so $\delta\ln(a)|_{\rho} < 0$ and we obtain a hot spot.
This was verified in~\cite{Wainwright:2013lea} for a specific choice of inflationary model.
However, when considering more complex models this simple intuition will not hold in general.
For example, in our two-field inflationary model~\eqref{eqn:potential_2field} the ``inflaton'' $\phi$ within each bubble has already begun to roll down the trough prior to the collision.
The effect of the collision, at least for the parameters used in this paper, is to pull $\phi$ back up the potential and to excite oscillations in $\sigma$.
Eventually the dynamical evolution of $\sigma$ produces oscillons, with the novel feature that the field $\phi$ is trapped at $\phi \approx 0$ in the interior of each oscillon.
If the net effect of the collision were to simply pull $\phi$ back up the potential, this would prolong the inflationary epoch and produce a positive $\delta\ln(a)|_\rho$.
However, the full dynamics creates additional contributions to the energy density (initially the oscillations in $\sigma$ and later the oscillons).
Thus, a detailed simulation including the effects of gravity is probably needed to ascertain the final sign of $\delta\ln(a)$, although it seems likely that positive values for the perturbation and thus cold spots can be generated. 

Finally, in our two-field model the inflationary epoch within the collision region restarts from a highly inhomogeneous state due to the population of oscillons.  
As long as these oscillons persist, they trap the inflaton field at the origin in regions of size $m_\sigma^{-1} = (\lambda_\sigma\sigma_0^2)^{-1/2}$.
Whether or not any remnant of this (non-vacuum) initial state persists is a very interesting question.
One possibility is that that once the oscillons decay, the pips of inflaton field that were held near the origin will be stretched in physical size as the universe expands and could end up sourcing curvature fluctuations $\zeta$.
As well, the field fluctuations in the collision regions will have initial conditions for the subsequent inflationary era that non-vacuum and highly nonGaussian.
If some trace of these inflaton perturbations survives, it would provide an example of spatially localized nonGaussian density fluctuations.

All of the effects speculated above will be rather strongly constrained by data, so a certain amount of tuning must be applied in order to construct models that produce signatures that are possibly observable but not yet ruled out by data.
As well, additional tuning is needed in order for the underlying theory to predict a nonnegligible number of potentially observable collisions for a typical observer.
Our purpose in this paper was simply to understand the dynamics of individual collisions so we will not touch on these model building issues here.
However, they must be addressed in order to assess the probability of seeing a given signal and to ultimately constrain the underlying theoretical model.

\subsection{Subhorizon Bubble Collisions}
Now let's consider the case when the bubble collisions occur in some first-order phase transition within our horizon.
Such a transition may have occurred in the past when the universe was extremely hot.
However, we can also envision a scenario where the phase transition is happening around the present time.
An interesting possibility would be for the field responsible for dark energy to undergo a first order phase transition.
As in the previous subsection, we restrict ourselves to signatures that result from collisions and the breaking of SO(2,1) symmetry,
although there are additional possibilities than are not of this type.
The discussion here in many ways mirrors the discussion for the superhorizon collisions, with the possible signals being more spatially homogeneous counterparts to the highly spatially localized effects discussed in section~\ref{sec:superhorizon_signatures}.

First consider the production of gravitational waves during a first order phase transition.
When the bubbles collide, gravitational waves are produced with wavelengths of order the typical size of the bubbles at collision.
This source of gravitational radiation is well known.
It is often studied using the envelope approximation, which neglects all of the field collision dynamics and treats the bubbles as if they were undergoing free expansion.
However, if the phase transition is not too rapid (so that the bubbles have a chance to grow before colliding), then we expect that the fracturing process explored in this paper will occur in individual collisions.
This will lead to an additional peak in the gravitational wave spectrum associated with the size of the unstable modes produced during the collision.
These two sources of gravitational waves mentioned above are a direct result of scalar field dynamics and are present even in vacuum.
However, in a high temperature phase transition the subsequent (turbulent) dynamics of the plasma may source additional gravitational waves.
In this case, the fracturing of the walls and long-lived oscillons may force turbulent motion and thus indirectly produce additional gravitational waves.

The production of oscillons during the phase transition also perturbs the expansion history of the universe by adding a component that dilutes as collisionless dust to the ambient cosmological constant (vacuum transition) or radiation bath (high temperature transition).
If the oscillons form a significant fraction of the post-transition energy density, this could lead to a temporary stage of matter domination in the early universe.
Since the linear growth of subhorizon adiabatic density perturbations is controlled by the expansion history, a temporary stage of matter domination could leave an observable remnant in large scale structure today.

\section{Applications to Other Scenarios}
\label{sec:extensions}
We have restricted ourselves in this paper to some very simple toy potentials that permit Coleman-deLuccia bubble nucleation.
However, the basic mechanism --- amplification of non SO(2,1) symmetric fluctuations --- likely plays a key role in many other scenarios.
For a small number of fields, the important feature of the potential appears to be the presence of a slightly asymmetric double-well structure in one of the field directions.
In this section we provide several examples from the recent literature where the asymmetric fluctuations were (we believe incorrectly) neglected.

A direct application is to the bubble baryogenesis scenario of~\cite{Cheung:2012im}.
This scenario takes a two-field model with slightly broken internal global $O(2)$ symmetry.
Baryon number is generated by the nucleation and expansion of bubbles, and also during bubble collisions.
The collisions considered by the authors of~\cite{Cheung:2012im} very closely resemble our results for the thin-walled bubbles, except that SO(2,1) symmetry was explicitly enforced.
It would be interesting to see how the net baryon number production is influenced by breaking of SO(2,1), as well as the extent to which baryon number is trapped within the oscillons.

Another possible application of these results is to boom and bust inflation~\cite{Brown:2008ea} and other models of flux discharge cascades driven by bubble nucleation and expansion in a compact extra dimension~\cite{Kleban:2011cs,D'Amico:2012sz}.
However, there are several caveats to a direct application of our results.
First of all, the bubble walls in these models can move at extremely relativistic speeds and are thus extremely Lorentz contracted in the center of mass frame for the collision.
Therefore, the duration of each collision is very short and the walls may simply pass through each other before the fracturing process has time to occur.
A second caveat is that the bubble walls in the flux cascade picture are D-branes instead of scalar field kinks.
Since the nonlinear interactions that lead to the breakup of the walls are sensitive to the details of the high energy completion of the theory, 
the nonlinear fracturing stage may be altered in the flux discharge case.

A final recent scenario appearing in the literature where non-symmetric fluctuations are likely to play a key role but were not considered is in~\cite{Ahlqvist:2013whn}.
Here, the authors study late-time violation of the free-passage approximation by displacing the field to very near local maximum of the potential in the collision region.
However, they make the assumption of planar symmetry and work in $1+1$ dimensions.
The nonplanar fluctuation modes will experience a tachyonic instability~\cite{Felder:2001kt,Felder:2000hj} that quickly invalidates the planar symmetry approximation.
In fact, the seeds of this are already in the author's work as their effect relies on precisely this tachyonic instability for the $k_\perp=0$ mode.

Beyond these obvious applications of the ideas presented in this paper,
there are many other scenarios in which corrections to SO(2,1) symmetry may be expected.
In this paper, we only considered collisions between bubbles formed from the same instanton.
Therefore, there were no topological constraints preventing the walls of the colliding bubbles from eventually annihilating each other.
A natural extension is to consider more complex potentials with many local minima.
Multiple instantons describing different decay channels for a single metastable minima can then exist, and collisions can occur between bubbles with different interiors. 
In this case, the field values in the bubble interiors enforce topological constraints on the collision dynamics.
Therefore, in general new domain walls interpolating between the bubble interiors will form in the collision region.
These newly formed domain walls can have their own internal dynamics which can again drive instabilities that lead to a breakdown of the SO(2,1) invariance.
This is the generalization of the oscillations of the planar shape mode in the double well potential studied in~\cite{ref:bbm2} 
to the case of bubble collisions.
Since the internal dynamics generally emit radiation, and this radiation will not in general respect the SO(2,1) symmetry, 
any signatures that may result from it again differ from the results of an SO(2,1) simulation.

\section{Conclusion}
\label{sec:conclusions_bub}
We performed fully three-dimensional simulations of bubble collisions in scalar field theories. 
Our treatment is novel because we include, for the first time, the effects of quantum fluctuations (in the semiclassical wave limit) on the dynamics.
Recent interest in this topic has been driven largely by false vacuum eternal inflation and early universe phase transitions.
In these instances, bubbles initially form through either quantum or thermal nucleation then subsequently undergo collisions.
In the case of bubbles nucleating in a false vacuum, previous studies have assumed that collisions between two such bubbles possesses an exact SO(2,1) symmetry.
This symmetry derives from the partial breaking of the SO(4) symmetry of the instanton solution.
However, it is important to keep in mind that the instanton only describes the most likely bubble to nucleate.
In reality, the actual shape of the nucleated bubble will be slightly deformed from the perfect instanton profile due to the quantum nature of the nucleation.
Even ignoring these deformations to the initial bubble shape, the bulk fluctuations in the ambient spacetime and inside the bubble are present after the nucleation of a bubble.
In fact, these subhorizon fluctuations inside the bubble must be present for inflation inside the bubble to seed density perturbations for the standard hot big bang.
In other applications, such as high temperature phase transitions, the nucleation of individual bubbles is again driven by coherent structures arising from stochastic fluctuations.
Hence in all these cases the exclusion of fluctuations, which is implicit in imposing exact SO(2,1) invariance, is \emph{not} consistent with the process of nucleation or creating a viable inflationary model.
Since the fluctuations do not obey the assumed SO(2,1) symmetry (or SO(2) symmetry in the case of thermal nucleation), it is important to test these assumptions.\footnote{As noted before, by this we mean that individual realization of the fluctuations do not preserve the symmetry, even if they do preserve the symmetry in a statistical sense.}

We studied collisions between pairs of nucleated bubbles in a variety of single-field and two-field potentials using highly accurate nonlinear three-dimensional lattice simulations.
A novel aspect of this investigation was the development of a pseudospectral approach for finding SO(4) symmetric instanton solutions, rather than the ubiquitous overshoot-undershoot method.
The accuracy of our instanton profiles are only limited by machine precision roundoff errors, and the procedure easily generalizes to multifield potentials.
Although we only considered the case of instantons in Minkowski space, our method can be trivially generalized to the case of a fixed background geometry,
and with a little extra effort to the case of a dynamical metric coupled to the scalar fields.

We studied two types of single-field potentials with radically different collision dynamics: double-well potentials with a broken $Z_2$ symmetry, and potentials with a single local minimum and a linear plateau which the field tunnels onto.
Under the assumption of SO(2,1) symmetry, collisions in the double-well potentials with mildly broken $Z_2$ symmetry cause the bubble walls to bounce off of each other many times during the collision.
These bounces produce an outgoing pattern of toroidal waves centered on location of the initial collision and expanding in the plane orthogonal to the collision axis.
When we include symmetry breaking fluctuations to model quantum effects, the fluctuations experience a strong instability as a result of the bouncing motion in the SO(2,1) background solution.
These fluctuations are quickly driven into the nonlinear regime, at which point a split into an SO(2,1) symmetric background and nonsymmetric fluctuations loses its utility.
A fully three-dimensional nonlinear description is required to describe the subsequent evolution.
The nonlinear dynamics causes the bubble walls and expanding torii to fracture into a network of localized blobs of field known as oscillons.
At this point, the SO(2,1) symmetry is spoiled completely and any dimensional reduction based on it will be a very poor approximation to the collision dynamics.
To the best of our knowledge, this is the first example where the three-dimensional nature of the problem plays an important role in the field dynamics for the collision between a pair of bubbles.\footnote{Easther et. al also performed three-dimensional bubble simulations~\cite{Easther:2009ft}, but they did not include fluctuations in their initial conditions and the effect that they discuss can be captured using 1+1-dimensional simulations.}
Meanwhile, in situations where the field tunnels out far away from a minimum of the potential (either because the $Z_2$ symmetry of the double-well is strongly broken or there is a long plateau), we found no evidence that the SO(2,1) breaking fluctuations experienced instabilities.\footnote{For the badly broken $Z_2$ double-well, this statement may be an artifact of our choice of potential, but for the plateau in the tunnel out region it should be generic.}
As a result, these situations can be well-approximated by dimensionally reduced simulations.

In the single field models we considered, it is difficult to have both an inflationary stage in the interior of the bubble while simultaneously having collisions that drive an instability in the symmetry breaking fluctuations.
This essentially follows from the fact that inflation requires a long region of the potential whose characteristic mass scale is much less than the Hubble constant.
In contrast, the bouncing of the walls and subsequent production of oscillons only occurs if the typical mass scale of the potential minimum is much greater than the Hubble constant.
For the case of identical bubbles, which was the focus of this paper, if the field repeatedly returns to the false vacuum in the collision region it must also repeatedly cross the portion of the potential that drives inflation.
Since the contraints of the curvature of the potential for slow-roll inflation and oscillon production are mutually exclusive, this suggests it is extremely difficult to both drive inflation within the bubble and produce oscillons in collisions for typical single field potentials.
To avoid this obstacle, we considered a very simple two-field model where we allowed one of the field directions to dominate the tunnelling and the other direction to drive slow-roll inflation.
Since current theoretical ideas suggest there are many effective degrees of freedom at the energy scales relevant for inflation,
the introduction of an additional field is quite natural from a top down perspective.
During a collision between two bubbles in the two-field model, the evolution in the tunnelling direction closely resembles that of the single-field collision in the double-well with mild $Z_2$ breaking.
As a result, oscillons are able to form in this direction.
The coupling between the tunnelling field and the inflaton direction then causes the ``inflaton'' field to be pulled back up the potential.
This entire process again badly breaks the SO(2,1) symmetry and requires more than a one-dimensional simulation to capture.
As well, the overall effect on the expansion history in the collision region is not necessarily to reduce the duration of inflation.
This is in contrast to the single field case with the field rolling down a plateau, where the collision simply displaces the field down the plateau.
Even in this relatively simple extension beyond single field models, the full collision dynamics is not well approximated by the theory of a single effective field.
Therefore some caution should be used when drawing conclusions about collisions in models with a large number of fields.
Since the fields now have many more directions to explore, both the tunnelling process and collision dynamics could change substantially in this limit.

Our simulations are not fully relativistic, so we are unable to track the subsequent evolution of the fields through the complete phase of inflation inside the bubble.
This is certainly an interesting question, but it is also a very challenging numerical problem that is much more difficult than the corresponding symmetry reduced (1+1)-dimensional problem.
However, since we can construct models where the time scale for oscillon formation is much less than the ambient Hubble time, 
our results set the initial conditions for the subsequent inflationary epoch within the collision region.
Since the SO(2,1) symmetry is broken so strongly in some cases, this question cannot be addressed within the framework of a symmetry reduced 1+1-dimensional problem.
In particular, previous signatures associated with bubble collisions have assumed an azimuthal symmetry will hold, which results from the assumption of SO(2,1) dynamics for the spacetime.

\acknowledgments
We would like thank Andrei Frolov and Belle Helen Burgess for useful discussions and comments.
This work was supported by the National Science and Engineering Research Council of Canada and the Canadian Institute for Advanced Research.
JB was partially supported by the European Research Council under the European Community's Seventh Framework Programme (FP7/2007-2013) / ERC grant agreement no 306478-CosmicDawn.
Computations were performed on CITA's Sunnyvale cluster.

\bibliography{bubble_collision_full}{99}

\end{document}